\documentclass[prb,amssymb,amsmath,twocolumn,showpacs]{revtex4}
\usepackage{amsfonts}
\usepackage{amsmath}
\usepackage{amssymb}
\usepackage[english]{babel}
\usepackage{fontenc}
\usepackage{graphicx}
\usepackage{color}
\usepackage{bbm}
\usepackage{natbib} 

\newcommand{\sgn}{\mathrm{sgn}}

\makeatletter
\newcommand{\ulinks}[3]{%
  \@mathmeasure\z@\displaystyle{#3}%
  \global\setbox\@ne\vbox to\ht\z@{}\dp\@ne\dp\z@
  \setbox\tw@\box\@ne
  \@mathmeasure4\displaystyle{\copy\tw@#1}%
  \@mathmeasure6\displaystyle{#3#2}%
  \dimen@-\wd6 \advance\dimen@\wd4 \advance\dimen@\wd\z@
  \hbox to\dimen@{}\mathop{\kern-\dimen@\box4\box6}%
}
\makeatother
\newcommand{\bp}{\begin{pmatrix}}
\newcommand{\ep}{\end{pmatrix}}

\newcommand{\measure}{\mathrm{d}}

\begin{document}
\title{Enhancement of Cooper pair splitting by multiple scattering}
\author{Martina Fl\"oser}

\affiliation{Institut NEEL,
  CNRS and Universit\'e Joseph Fourier, BP 166,
  F-38042 Grenoble Cedex 9, France}

\author{Denis Feinberg}

\affiliation{Institut NEEL,
  CNRS and Universit\'e Joseph Fourier, BP 166,
  F-38042 Grenoble Cedex 9, France}

\author{R\'egis M\'elin}
\email{Regis.Melin@grenoble.cnrs.fr}
\affiliation{Institut NEEL,
  CNRS and Universit\'e Joseph Fourier, BP 166,
  F-38042 Grenoble Cedex 9, France}
\begin{abstract} 
In three-terminal NSN hybrid structures the influence of additional barriers on the nonlocal conductance and on current cross-correlations is studied within a scattering theory. In metallic systems with additional barriers and phase averaging, which simulate disordered regions, local processes can be enhanced by reflectionless tunneling but this mechanism has little influence on nonlocal processes and on current cross-correlations. Therefore Cooper pair splitting cannot be enhanced by reflectionless tunneling. On the contrary, in ballistic systems, additional barriers lead to Fabry-Perot resonances and allow to separate the different contributions to the conductance and to the current cross-correlations. In particular, crossed Andreev processes can be selectively enhanced by tuning the length or the chemical potential of the interbarrier region.
\end{abstract}
\maketitle

\section{Introduction}
Transport in normal metal-superconductor-normal metal (N$_a$SN$_b$) hybrid three-terminal nanostructures offers a special interest, owing to the suggestion of producing split pairs of spin-entangled electrons from a superconductor \cite{recher,lesovikmartin}. This is possible when the size of the region separating the N$_a$S and the N$_b$S interface becomes comparable to the superconducting coherence length. This circumstance indeed allows coherent processes involving two quasi-particles, each simultaneously crossing one of the two interfaces \cite{allsopp,byers,torres,deutscher,falci,melin1,melin2}. Much effort has been devoted to the theoretical understanding and to the experimental observation of such a Cooper pair splitting effect. In a transport experiment where electrons are difficult to measure one by one --- contrarily to the similar production of entangled photons ---, one should extract the relevant information from steady transport measurement, e.\,g. the current-voltage characteristics (conductance) and the cumulants of the current fluctuations (non-equilibrium current noise \cite{antibunching,bignon} and its counting statistics \cite{fazio}). In practice, the conductance and the second-order cumulant (the shot noise and the current cross-correlations between the two current terminals N$_a$, N$_b$) are the quantities to be extracted from experiments. Indeed, due to Fermi statistics, the ``partition`` noise correlations at a three-terminal crossing of normal metal contacts are negative~\cite{antibunching}, manifesting the antibunching properties of individual electrons. If instead one contact is made superconducting, the cross-correlations may become positive, indicating the splitting of Cooper pairs~\cite{torres}.

There are two basic nonlocal processes occurring at the double NS interface: Crossed Andreev Reflection (CAR) which alone leads to Cooper pairs splitting into separated electrons bearing opposite spins (for a spin singlet superconductor), and Elastic Cotunneling (EC) which alone leads to (spin-conserving) quasi-particle transmission between the normal contacts, across the superconducting gap   \cite{falci}. For tunnel contacts, at lowest order in the barrier transparencies, those two processes are simply related to the conductance, leading to positive (resp. negative) conductance and current cross-correlations for CAR (resp. EC) processes. Indeed Bignon \textit{et\,al.}~\cite{bignon} showed that for tunnel contacts the linear dependence of the current cross-correlation on the voltages applied to the contacts N$_a$, N$_b$ allows to separately track the amplitudes of CAR and EC. Due to the expected compensation of the opposite CAR and EC conductance components at low transparencies, ferromagnetic contacts are required to detect CAR and EC from the conductance with tunnel contacts \cite{deutscher,falci,sanchez}. Yet, such polarizations are not easily achievable, moreover, if one is interested in producing spin-entangled electrons in a nonlocal singlet state, one should of course not spin polarize the contacts.  

Yet, as regards experiments, the situation for (extended) tunnel barriers
looks more complicated than given by a simple tunnel model
\cite{DelftExpt}. In addition, zero-frequency noise measurements involve weak
signals and therefore they require more transparent interfaces, so that the arguments based on a tunnel model cannot be directly applied. For more transparent contacts, it has been found theoretically \cite{melin2,duhotmelin,brataas,kalenkovzaikin,FFM2010} that the nonlocal conductance is dominated by EC thus is negative, which leaves the current cross-correlations as the only possible proof of Cooper pair splitting processes, provided one controls the voltages on both contacts. Contrarily to conductance measurements \cite{exptconductance}, cross-correlations have led to few experimental results \cite{exptcrossnoise,das2012}. At the theoretical level, the dependence of the cross-correlations on the contact transparency is not yet fully understood. Further analysis is required if one wishes to extract information from the sign and the amplitude of both the nonlocal conductance and the current cross-correlations in a system where the contacts are non-ferromagnetic metals and the interface transparencies are rather large. 

In view of the current experiments on metallic structures, the main question is therefore: Can the cross-correlations be positive, and if the answer is yes, is this a signature of Cooper pair splitting ? Previous work on a NSN structure~\cite{melinbenjamin} showed that the cross-correlations can indeed be positive at large transparencies, although the nonlocal conductance is negative. The origin of this somewhat surprising result was not fully elucidated. Further work~\cite{FFM2010} showed that the sign of the cross-correlations indeed changes with the transparency of the interfaces, being positive at low transparency, negative at intermediate transparency and positive again at high transparency. While the positive sign at low transparency is clearly ascribed to Cooper pair splitting, it was shown that the positive sign at high transparency should not be interpreted in the same way. Indeed, at high transparency, CAR processes do not dominate either in the conductance or in the noise. Instead, the positive cross-correlations should be ascribed to higher order processes, in particular those involving local Andreev reflection (AR) on one side, and the opposite process on the other side, a process equivalent to exchanging a pair of electrons between the two normal contacts.

These results show that transparent interfaces are detrimental to Cooper pair splitting. A possible alternative strategy to boost Cooper pair splitting is to take advantage of multiple scattering to enhance the CAR process. Indeed, at a single NS interface, it was shown that disorder in the N region, or multiple scattering at a clean NN$_l$S double interface, can strongly enhance Andreev reflection, a mechanism nicknamed ''reflectionless tunneling'' \cite{reflectionlesstunnexpt,reflectionlesstunntheo,MB1994}. With a disordered N' region, it holds below the Thouless energy, where the mutual dephasing of electrons and Andreev-reflected holes is negligible. In the case of a clean double NN$_l$S interface, maximum Andreev transmission is obtained by balancing the transparencies T$_{nn}$ and T$_{ns}$ of the NN$_l$ and N$_l$S interfaces, such that the N$_l$ region acts as a resonant cavity. Melsen and Beenakker \cite{MB1994} performed an average on the modes inside N$_l$ in order to mimick a disordered region. One can wonder whether a similar mechanism could enhance nonlocal Andreev reflection e.g. boost CAR compared to AR and EC. For this purpose we consider in this work a set-up N$_a$N$_l$SN$_r$N$_b$ with a quadruple interface. This system can be either considered as a modelization of a true one-dimensional hybrid structure with clean (ballistic) normal and superconducting elements separated by tunnel barriers, for instance using carbon nanotubes or semiconducting nanowires. Or it can mimick a NSN system where the N metals are disordered close to the S interface. A scattering theory is performed in a one-dimensional geometry, varying the transparencies of the barriers and the width of the superconductor. 

One important result of this work is the following : If averaging independently the modes in the left and the right regions N$_l$, no boosting of the CAR process is obtained. Indeed, ``reflectionless tunneling`` is a quantum coherent process which demands that the Andreev-reflected hole retrace the path of the electron, by scattering on the same impurities. On the contrary, with nonlocal Andreev reflection, the electron and the transmitted hole sample different disorders and no coherence is obtained. Thus, the common « reflectionless tunneling » mechanism cannot enhance Cooper pair splitting in such a disordered one-dimensional geometry. Interestingly, we show that if on the contrary the N$_l$ regions can be tuned at a Fabry-Perot resonance for the specific electron and Andreev-transmitted hole wavevectors, then CAR can be efficiently enhanced, moreover CAR and EC processes can be filtered at will by tuning the energy or the Fermi wavevector of both N$_l$ region. The feasibility of such a proposal is discussed at the end.

Section~\ref{sec:paperNSN:model} presents the model and section~\ref{paperNNSNN:sec:comp} the scattering theory of the N$_a$N$_l$SN$_r$N$_b$ system. Section~\ref{paperNNSNN:sec:AS} provides the results obtained by averaging over channels in N$_l$ and N$_r$, in the spirit of Ref.~\onlinecite{MB1994}. Section~\ref{sec:NSN:Ballistic} describes transport in the case where N$_l$ and N$_r$ are clean metallic ``cavities''. Section~\ref{NSNpaper:sec:ExpObs} concludes by discussing a possible implementation of a controllable Cooper pair splitter.
\section{The model\label{sec:paperNSN:model}}
\begin{figure}
\includegraphics[width=0.45\textwidth]{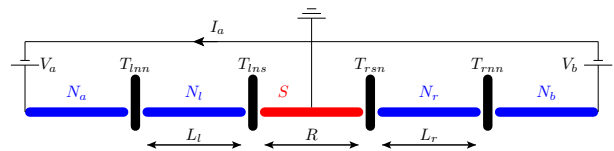}
 \caption{\label{paperNNSNN:fig:model} Schematic of the model.}
\end{figure}

We study a one-dimensional model of a symmetrical three-terminal normal
metal-superconductor-normal metal hybrid structure depicted in
Fig.~\ref{paperNNSNN:fig:model}. The central superconducting electrode is
grounded, the normal terminals can be biased with voltages $V_a$ and $V_b$.
The length $R$ of the superconducting electrode is comparable to the
superconducting coherence length, which makes the observation of
nonlocal effects possible. The interfaces between the normal metal and the superconducting electrodes are modeled by barriers with transparencies $T_{lns}$ and $T_{rsn}$. In both normal metal electrodes there is an additional barrier at a distance $L_l$ (respectivly $L_r$) from the normal metal superconductor interface with transparency $T_{lnn}$ (respectivly $T_{rnn}$).\\

The system can be
described by a $4\times4$ scattering matrix $s_{ij}^{\alpha\beta}$ where Latin 
indices run over the normal electrodes $a$ and $b$ and Greek indices over
electrons $e$ and holes $h$. For energies lower than 
the superconducting gap, there are no single electron or hole states in the bulk of the
superconducting
electrode. Consequently, there is no scattering of electrons or holes in or out of
the reservoir in the superconducting electrode and the latter
does not appear in the scattering matrix. However, the scattering theory implicitly assumes that the superconductor is a reservoir of Cooper pairs. It is thus essential that the
superconducting electrode is grounded. The transformation of quasi-particles into
Cooper pairs is taken into account by the correlation length $\xi$
which sets the scale of the damping of the electron and hole wavefunctions in the superconductor.\\

The elements of the scattering matrix are evaluated from the BTK
approach\cite{BTK1982} (see Appendix~\ref{paperNSN:app:BTK}). As the energy dependence of the calculated quantities is relevant, we do not use the Andreev approximation valid in the limit of zero energy where the electron and hole wavevectors are set to the Fermi wavelength, but instead keep the full expressions for the wave vectors.
The calculation of the average current and current cross-correlations rely on the formulas obtained by
Anantram and Datta in Ref.~\onlinecite{AD1996}:
\begin{align}
&I_i=\frac{e}{h}\sum_{k\in \{a,b\}}\sum_{\alpha,\beta\in\{e,h\}} \sgn(\alpha)\label{paperNSN:eq:current}\\
&\times\int dE\phantom{0} \left[\delta_{ik}\delta_{\alpha\beta}-\left|s_{ik}^{\alpha\beta}\right|^2\right]f_{k\beta}(E)\notag\\
&S_{ij}=\frac{2e^2}{h}\sum_{k,l\in \{a,b\}}\sum_{\alpha,\beta,\gamma,\delta\in\{e,h\}} \sgn(\alpha)\sgn(\beta)\\
&\times\int dE\phantom{0}A_{k\gamma,l\delta}(i\alpha,E)A_{l\delta,k\gamma}(i\alpha,E)f_{k\gamma}(E)\left[1-f_{l\delta}(E)\right]\notag
\end{align}
with $ A_{k\gamma,l\delta}(i\alpha,E)=\delta_{ik}\delta_{il}\delta_{\alpha\gamma}\delta_{\alpha\delta}-s_{ik}^{\alpha\gamma\dagger}s_{il}^{\alpha\delta}$, $\sgn(\alpha=e)=1$ $\sgn(\alpha=h)=-1$, 
$f_{i\alpha}$ the occupancy factors for the electron and hole states in
electrode i, given by the Fermi function where the chemical potential are the
applied voltages 
$f_{ie}(E)=\left[1+\exp\left(\frac{E-V_i}{k_BT}\right)\right]^{-1}\xrightarrow[T\to0]{}\theta(-E+V_i)$, 
$
f_{ih}(E)=\left[1+\exp\left(\frac{E+V_i}{k_BT}\right)\right]^{-1}\xrightarrow[
T\to0]{}\theta(-E-V_i)$.\\

 In this one-dimensional model, both current and noise are highly sensitive to the distances $L_l$, $R$, $L_r$ between the barriers: They
oscillate as a function of these distances with a period equal to the Fermi wavelength
$\lambda_F\ll L_l, R, L_r$. In a ballistic system, the multiple barriers act like a Fabry-Perot interferometer and the interference effects are studied in section~\ref{sec:NSN:Ballistic}.  
In a higher-dimensional system with more than one transmission mode,
the oscillations in the different modes are independent and are thus averaged
out. Multidimensional behavior can be simulated qualitatively with a
one-dimensional system by averaging all quantities over one oscillation
period:
\begin{align}
\label{PaperNSN:eq:av}
&\overline{X}(L_l, R, L_r)\\
&= \frac{1}{\lambda_F^3}
\int_{L_l-\frac{\lambda_F}{2}}^{L_l+\frac{\lambda_F}{2}} dl_l\int_{R-\frac{\lambda_F}{2}}^{R+\frac{\lambda_F}{2}} dr \int_{L_r-\frac{\lambda_F}{2}}^{L_r+\frac{\lambda_F}{2}} dr\kern2pt X(l_l,r,l_r).\notag
\end{align}
This procedure is appropriate to describe metallic systems. These averaged quantities are studied in section \ref{paperNNSNN:sec:AS}.

\section{Components of the Differential Conductance and the Differential Current Cross-Correlations\label{paperNNSNN:sec:comp}}
An electron, arriving from one of the normal metal reservoirs at the interface
to
the superconductor, can have four different destinies: 
It can be reflected as an electron (normal reflection (NR)), or reflected as a
hole (Andreev reflection (AR)), or transmitted as an electron (elastic
cotunneling (EC)) or transmitted as a hole (crossed Andreev reflection
(CAR)), and similarly for
holes.
The corresponding elements of the scattering matrix are for
NR: $s_{aa}^{ee}$, $s_{aa}^{hh}$, $s_{bb}^{ee}$, $s_{bb}^{hh}$,
AR: $s_{aa}^{eh}$, $s_{aa}^{he}$, $s_{bb}^{eh}$,
$s_{bb}^{he}$,
EC: $s_{ab}^{ee}$, $s_{ab}^{hh}$, $s_{ba}^{ee}$, $s_{ba}^{hh}$,
and CAR: $s_{ab}^{eh}$, $s_{ab}^{he}$, $s_{ba}^{eh}$,
$s_{ba}^{he}$.\\

The current in electrode $N_a$ given by Eq.~\eqref{paperNSN:eq:current} can
naturally be divided into AR, CAR and EC contributions (to calculate this
expression the unitarity of the scattering matrix has been used):
\begin{widetext}
\begin{align}
 I_a=\frac{|e|}{h}\int \measure E &\underbrace{\left[\left(|s_{aa}^{eh}(E)|^2+|s_{aa}^{he}(E)|^2\right)(f_{ae}(E) - f_{ah}(E))\right.}_{\text{local Andreev reflection}}\notag\\
+&\underbrace{|s_{ab}^{ee}(E)|^2(f_{ae}(E) - f_{be}(E))+|s_{ab}^{hh}(E)|^2(f_{bh}(E)-f_{ah}(E))}_{\text{elastic cotunneling}}\notag\\
+&\underbrace{\left.|s_{ab}^{eh}(E)|^2(f_{ae}(E) - f_{bh}(E))+|s_{ab}^{he}(E)|^2(f_{be}(E)-f_{ah}(E))\right]}_{\text{crossed Andreev reflection}}.
\end{align}
In the following, we focus on i) the differential  conductance in the symmetrical case where $V_a=V_b=V$ and the current $I_a$ is differentiated with respect to $V$, and ii) the differential  nonlocal conductance in the asymmetrical case where $V_a=0$ and the current $I_a$ is differentiated with respect to $V_b$. In the zero temperature limit, only the nonlocal processes, CAR and EC, contribute to the nonlocal conductance:
\begin{align}
 \left.\frac{\partial I_a}{\partial V_b}\right|_{V_a=0}
=\underbrace{-\frac{e^2}{h}\left[
|s_{ab}^{ee}(|e|V_b)|^2+|s_{ab}^{hh}(-|e|V_b)|^2\right]}_{\text{elastic cotunneling}}
+\underbrace{\frac{e^2}{h}\left[|s_{ab}^{eh}(-|e|V_b)|^2+|s_{ab}^{he}(|e|V_b)|^2\right]}_{\text{crossed Andreev reflection}},
\end{align}
 while the symmetric case contains local Andreev reflection and crossed Andreev reflection:
\begin{align}
\left.\frac{\partial I_a}{\partial V}\right|_{V_a=V_b=V}
=&\underbrace{\frac{e^2}{h}\left[\left(|s_{aa}^{eh}(|e|V)|^2+|s_{aa}^{he}(|e|V)|^2\right)+\left(|s_{aa}^{eh}(-|e|V)|^2+|s_{aa}^{he}(-|e|V)|^2\right)\right]}_{\text{local Andreev reflection}}\notag \\
+&\underbrace{\frac{e^2}{h}\left[\left(|s_{ab}^{eh}(|e|V)|^2+|s_{ab}^{he}(|e|V)|^2\right)+\left(|s_{ab}^{eh}(-|e|V)|^2+|s_{ab}^{he}(-|e|V)|^2\right)\right]}_{\text{crossed Andreev reflection}}
\end{align}
Let us now perform a similar analysis for the current cross-correlations. We
study only the zero temperature limit, where
$f_{k\gamma}(E)[1-f_{l\delta}(E)]$ is zero if
$k=l$ and $\gamma=\delta$ and the current 
cross-correlations are:
\begin{equation}
 S_{ab}(T=0)=\frac{2e^2}{h}\sum_{k,l\in \{a,b\}}\sum_{\alpha,\beta,\gamma,\delta\in\{e,h\}}\sgn(\alpha)\sgn(\beta)\int dE 
s_{ak}^{\alpha\gamma\dagger}s_{al}^{\alpha\delta}s_{bl}^{\beta\delta\dagger} s_{bk}^{\beta\gamma}f_{k\gamma}(E)[1-f_{l\delta}(E)]
\end{equation}
\end{widetext}
Every summand in $S_{ab}$ contains the product of four elements of the
scattering matrix. As pointed out in Refs. \onlinecite{ML1992, Buettiker1990}, in
difference to the situation for the current, it is impossible to combine those
matrix elements to absolute squares.
Let us now try to classify the contributions of the noise as we did above for
the current. We find that no summand consists
of only one kind of elements of the scattering matrix. Every element consists of
two local elements (NR or AR ) and two nonlocal elements (CAR or EC). Either
the two local elements and the two nonlocal elements are identical, that gives
the components EC-NR, 
CAR-NR, EC-AR, CAR-AR, or all four matrix elements belong to different
categories and we will call these summands MIXED. Sometimes, it is useful to divide MIXED further as a function of its voltage dependence (see Appendix~\ref{paperNNSNN:app:Noise}).
As the formulas for the current cross correlations are lengthy, they are relegated into appendix~\ref{paperNNSNN:app:Noise}.

For the interpretation of current cross-correlations, the global sign plays an important role. For positive applied bias voltages, the differential current cross-correlations carry the same sign as the current cross-correlations. For negative applied voltages current, cross-correlations and differential current cross-correlations have opposite signs. To avoid confusion, we only show pictures of the differential current cross-correlations calculated for positive bias voltages (and thus negative energies $E=-|e|V$). Due to the electron-hole symmetry of the model, differential current cross-correlations calculated for negative bias voltages are up to a global sign identical to the ones calculated at positive bias voltage. For small bias voltages, current cross-correlations depend linearly on voltage. Thus, current cross-correlations and differential current cross-correlations show the same qualitative behavior if studied as a function of the interface transparency or the distance between barriers.

 In Ref.~\onlinecite{FFM2010}, we performed a similar analysis of current
cross-correlations in terms of
Green's functions. For the relations between these two
classifications see Appendix~\ref{PaperNSN:app:Green}.\\ 
Bignon \textit{et\,al.}~\cite{bignon} have studied current cross-correlations
in the tunneling limit. They find that noise measurements in the
tunneling limit can give access to the CAR and EC contribution of the current.
We have just seen, that at least two processes are
involved in every component of noise, but the contributions of noise they calculate
fall into the categories EC-NR and CAR-NR. In the tunneling limit the NR
contribution is so close to one, that it can be neglected and what remains is
very similar to the current contributions.
\section{The Averaged System \label{paperNNSNN:sec:AS}}
\subsection{Positive Cross-Correlations without CAR}
In fermionic systems without superconducting electrodes, current
cross-correlations are, as a consequence of the Fermi-Dirac statistics, always
negative~\cite{antibunching, AD1996}. In superconductors, on the contrary, the coupling of electrons
into Cooper pairs makes positive cross-correlations possible \cite{torres}. If a CAR process is interpreted as the splitting of a Cooper pair into two electrons
leaving the superconductor in different electrodes, positive cross-correlations
are its logical consequence. However, the CAR process is not the only one
which can lead to positive cross-correlations. Let us investigate in more detail the influence of the different processes on the
current cross-correlations in a NSN-system, i.\,e. in a system without the additional barriers in the normal conducting electrodes.

\begin{figure}
 \includegraphics[width=\columnwidth]{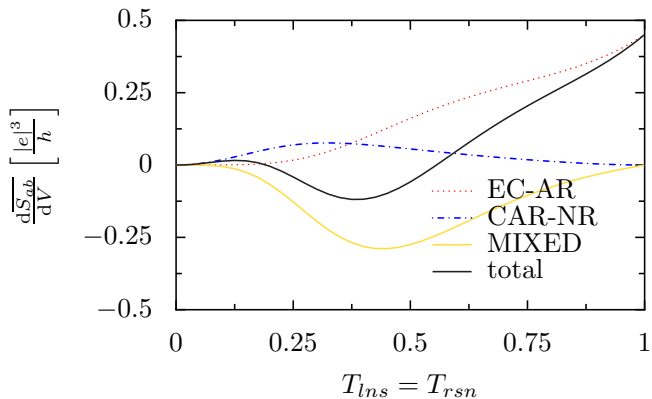}
\caption{\label{paperNSN:fig:ecnr}Averaged differential current cross-correlations for a symmetrical biased
 ($V=V_a=V_b\ll\Delta/|e|$) NSN-system as a function of the transparency of the interfaces $T_{lns}=T_{rsn}$. The positive cross-correlations at high interface transparency are due to the EC-AR process, represented by a dotted line.}
\end{figure}
The black line in Figure~\ref{paperNSN:fig:ecnr} shows the averaged differential current cross-correlations for symmetric bias
($V=V_a=V_b$). The total current cross-correlations have already been published in Ref.~\onlinecite{FFM2010}, but here, Figure~\ref{paperNSN:fig:ecnr} shows in addition the different parts which contribute to the total current cross-correlations.
The total cross-correlations are positive for high interface 
transparencies and for low interface transparencies. As we have already argued
in Ref.~\onlinecite{FFM2010}, the positive cross-correlations at high interface transparencies are not due to CAR.
Indeed, at very high transparencies, only processes which conserve momentum can occur, since
there are no barriers which can absorb momentum.
CAR processes do not conserve momentum: if e.g. an electron arrives from the
left hand side carrying momentum $k_F$, the hole that leaves at the right hand
side carries momentum $-k_F$. As the cross-correlations do not tend to zero
even for very high transparencies, they cannot be due do CAR.
Indeed, if we plot the different components of the noise introduced in the last section separately, we see that positive current cross-correlations at low interface transparencies are a consequence of a large CAR-NR component and therefore a consequence of CAR processes. But the positive current cross-correlations at high interface transparencies have a different origin: a large positive EC-AR contribution. 

We can put the contributions to the current cross-correlations into two categories with respect to their sign, which is independent of the interface transparency. EC-NR, CAR-AR, MIXED2 and MIXED4 carry a negative sign, CAR-NR, EC-AR, MIXED1 and MIXED3 carry a positive sign. The current can either be carried by electrons $I^e$ or by holes $I^h$. The sign of the different contributions to the current cross-correlations depends on whether only currents of the same carrier type are correlated~\cite{AD1996} ($\langle \Delta\hat I_a^e\Delta\hat I_b^e\rangle$+$\langle \Delta\hat I_a^h\Delta\hat I_b^h\rangle+a\leftrightarrow b$), which is the case for EC-NR, CAR-AR, MIXED2 and MIXED4 and leads to a negative sign; or whether electron currents are correlated with hole currents ($\langle \Delta\hat I_a^e\Delta\hat I_b^h\rangle$+$\langle \Delta\hat I_a^h\Delta\hat I_b^e\rangle+a\leftrightarrow b$), which is the case for CAR-NR, EC-AR, MIXED1 and MIXED3 and leads to a positive sign. In purely normal conducting systems, the electron and hole currents are uncorrelated, only correlations of the same carrier type contribute to the current cross-correlation and lead to a negative sign. The sign of the total current cross-correlations is a consequence of the relative strength of the different parts of the current cross-correlations, which depends on the interface transparency.
\subsection{Multiple Barriers}
In the last paragraph, we learnt that positive cross-correlations due to CAR can only be found in the tunneling regime.
But in the tunnel regime the signals are quite weak. 
The conductance over an NS-tunnel junction can be larger for a ``dirty`` normal conductor containing a large number of non-magnetic impurities, where transport is diffusive, than for a clean normal conductor, where transport is more ballistic, by an effect called reflectionless tunneling \cite{reflectionlesstunnexpt,reflectionlesstunntheo}.
It is thus natural to ask if a similar effect could also enhance conductance and current cross-correlations in a three terminal NSN-structure.
To answer this question, we use the model of Melsen and Beenakker \cite{MB1994} where the disordered normal conductor is replaced by a normal conductor with an additional tunnel barrier leading to an NNS structure.
Duhot and M\'elin \cite{duhotmelin} have studied the influence of additional barriers
on the nonlocal conductance in three terminal NSN-structures. They find
that two symmetric additional barriers enhance the nonlocal conductance.

First, let us get a deeper understanding of their result by
calculating the AR, CAR and EC components of the current separately.
Afterwards, we will study the influence of additional barriers on the current
cross-correlations.\\
\begin{figure}
a)\hspace{-0.5cm}\includegraphics[width=\columnwidth]{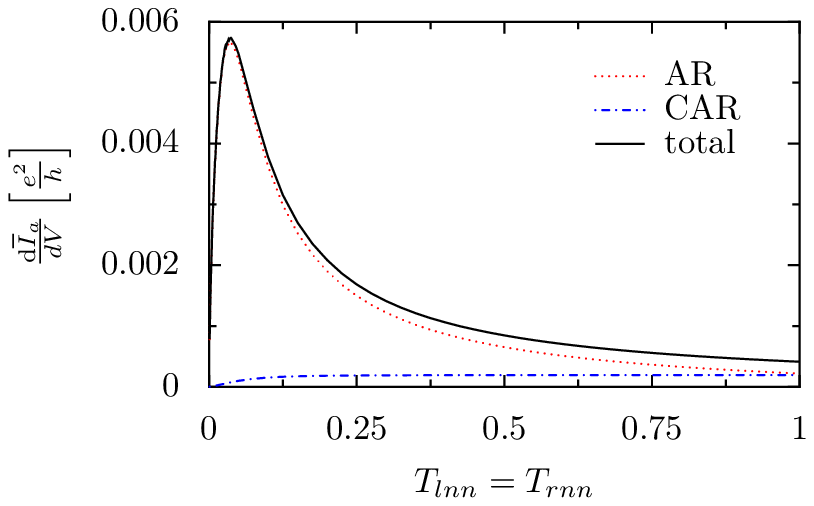}
b)\hspace{-0.5cm}\includegraphics[width=\columnwidth]{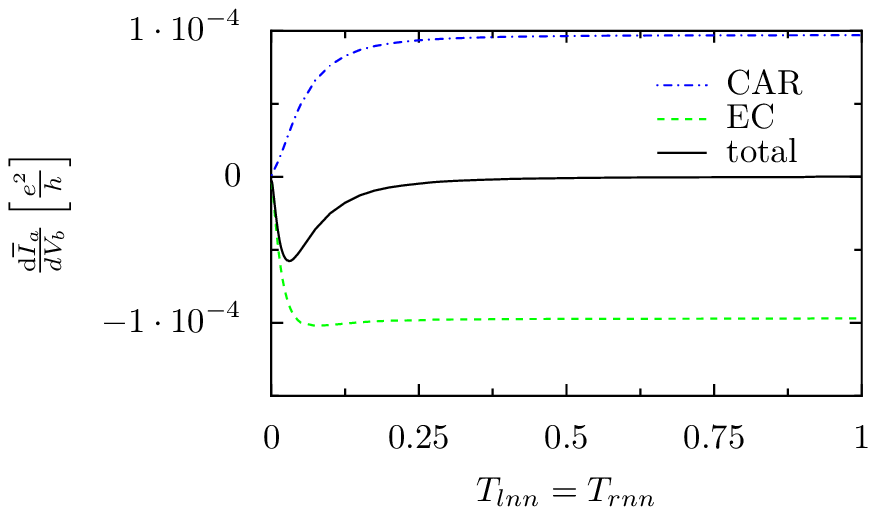}
\caption{\label{paperNNSNN:fig:AvCond}Averaged differential conductance in the limit of zero energy in a) the
symmetrical bias situation $V_a=V_b\ll\Delta/|e|$ and b) 
in the asymmetrical case $V_a=0$,
$V_b\ll\Delta/|e|$ for a superconducting electrode
much shorter than the coherence length ($R=0.25\xi$) as a function of the
transparencies of the additional
barriers $T_{lnn}=T_{rnn}$. The barriers next to the superconductor are in the
tunnel regime ($T_{lns}=T_{rsn}=0.01$).}
\end{figure}
\begin{figure}
a)\hspace{-0.5cm}\includegraphics[width=\columnwidth]{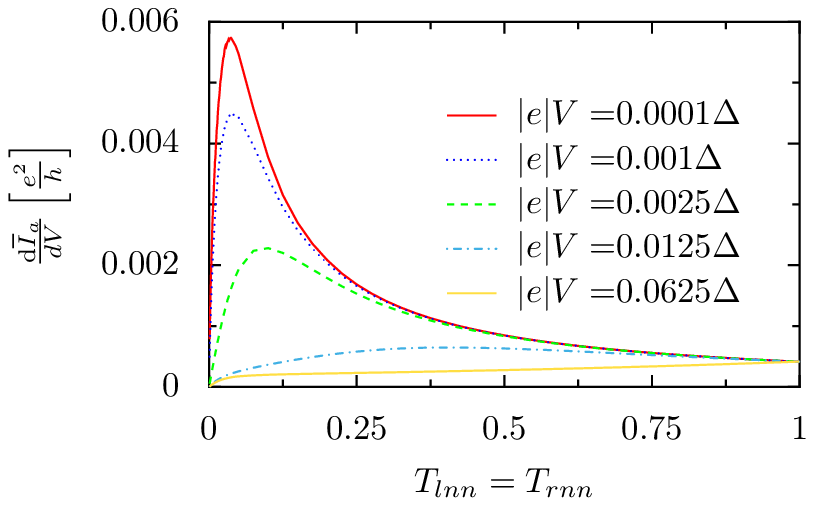}
b)\hspace{-0.5cm}\includegraphics[width=\columnwidth]{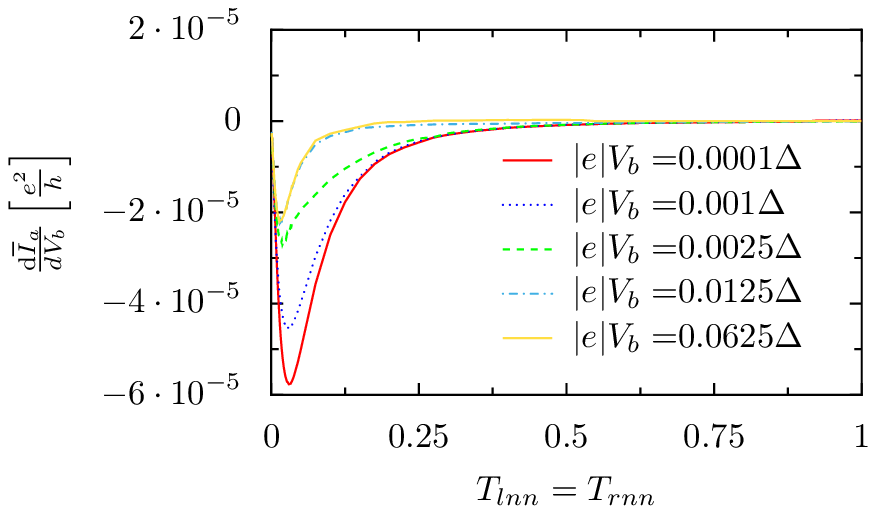}
\caption{\label{paperNSN:fig:EnDepAvCond} Total averaged differential conductance, as in Figure~\ref{paperNNSNN:fig:AvCond} but at different energies.}
\end{figure}
Figure~\ref{paperNNSNN:fig:AvCond} shows the averaged conductivity in the
symmetrical bias situation $V_a=V_b\ll\Delta/|e|$ and 
in the asymmetrical voltage case $V_a=0$,
$V_b\ll\Delta/|e|$ for a superconducting electrode
much shorter than the coherence length ($R=0.25\xi$). The sum of the AR, CAR and
EC components, traced in black, 
features in both cases an extremum. Yet, looking at the behavior of there
components, we see that they arise from different mechanisms. 
Let us first have a look at the symmetrically biased case. Without the additional barriers, i.e. in the limit $T_{lnn}=T_{rnn}\to1$, 
the contributions of AR and CAR are similar in magnitude. The EC component is
completely suppressed, since it is proportional
to the difference of the applied voltages. The introduction of two additional
barriers increases the AR component by about a 
factor 30. The shape of the curves is similar to the one of the NNS structure derived analytically by Melsen and Beenakker \cite{MB1994}, which
can be recovered exactly by increasing the
length of the superconducting electrode far beyond the coherence length. However, the CAR
curve stays almost constant
over a wide range of values of barrier strength of the additional barriers, and it eventually
vanishes when the transparencies
go to zero.

In the asymmetrical voltage case $V_a=0$, the AR component
is zero as it is proportional to the 
local voltage $V_a$. Like in the first case, the additional barriers have little influence on the CAR component, 
except for the fact that it tends to zero for vanishing transparency. Over a wide range of 
barrier strength values, the
EC component is identical in amplitude, but opposite in sign to the CAR
component. For small $T_{lnn}=T_{rnn}$ values the EC
component displays a small extremum, but it is much less pronounced than the
maximum of the AR component of the 
first case. In the limit $T_{lnn}=T_{rnn}\to1$, the EC component tends more slowly to zero than the CAR component which as a consequence has a minimum in the total conductance. The fact that the conductance maxima in the symmetrical bias case and in the asymmetrical bias case have different origins can also be illustrated by studying their energy dependence, depicted in Figure~\ref{paperNSN:fig:EnDepAvCond}: The enhancement of the AR component of the conductance in the symmetrical biased case disappears completely with increasing bias voltage. The extremum of the conductance in the asymmetrical biased case decreases slightly with increasing bias voltage, but only up to a certain voltage value, then it saturates.

\begin{figure}
\includegraphics[width=\columnwidth]{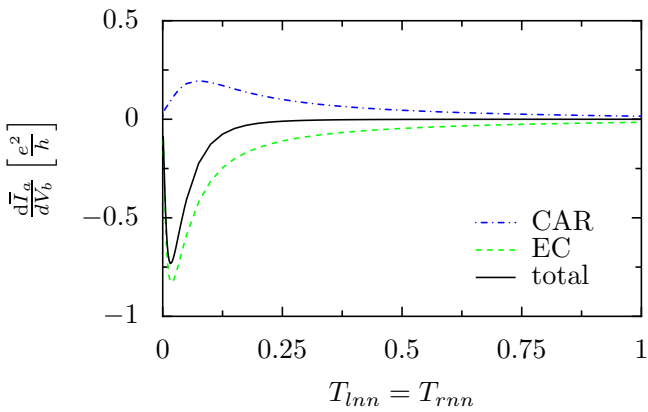}
\caption{\label{paperNSN:fig:Coupled}Gedankenexperiment with coupled
integrals $T_{lns}=T_{rsn}=0.01$, $R=0.25\xi$: Now, also the EC and the CAR component are enhanced by reflectionless tunneling.}
\end{figure}

Why is the AR component enhanced by the additional barriers, but not the EC or CAR components? Reflectionless tunneling
is believed to occur because the electrons and holes get localized between the double barrier and have therefore a 
higher probability to enter the superconductor. There is a double barrier on the
right and on the left hand side, so this localization should also happen to the particles involved in EC or CAR
processes. The answer is, that in the AR cases
the incoming electron and the leaving hole see the same environment. In the EC
and CAR process, the incoming particle
sees the environment on one side of the superconductor and the leaving particle
the environment on the other side. The energy dependence of the conductance enhancement of the AR component demonstrates that reflectionless tunneling can only occur if the outgoing hole can trace back the way of the incoming electron. At low bias voltages where the electron and the hole have nearly the same energy and wavevector, reflectionless tunneling occurs. At higher bias voltage, electrons and holes have different wavevectors and the reflectionless tunneling peak disappears.
The integrals over the phases between the additional barriers on the left- and on the right-hand side have, 
of course, been taken independently. There is no reason to think that the channel mixing, which is emulated by the 
integrals, on the left- and on the right-hand side are coupled.
To verify this scenario, let us couple the two
integrals in an gedankenexperiment. We set the distance $L_l$ between the two left-hand side barriers to be equal to the distance $L_r$ between the two right-hand side barriers and perform only one integral over $L=L_l=L_r$.  The result is shown in 
Figure~\ref{paperNSN:fig:Coupled}. Now, the CAR and the EC component are also
enhanced. Still, the effect on the EC component is 
larger and EC dominates the nonlocal conductance. \\

\begin{figure}
a)\hspace{-0.5cm}\includegraphics[width=\columnwidth]{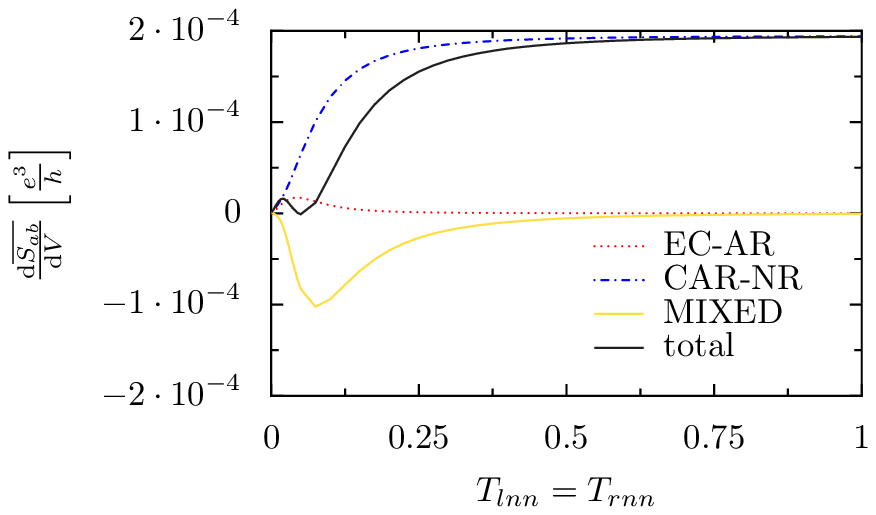}\\
b)\hspace{-0.5cm}\includegraphics[width=\columnwidth]{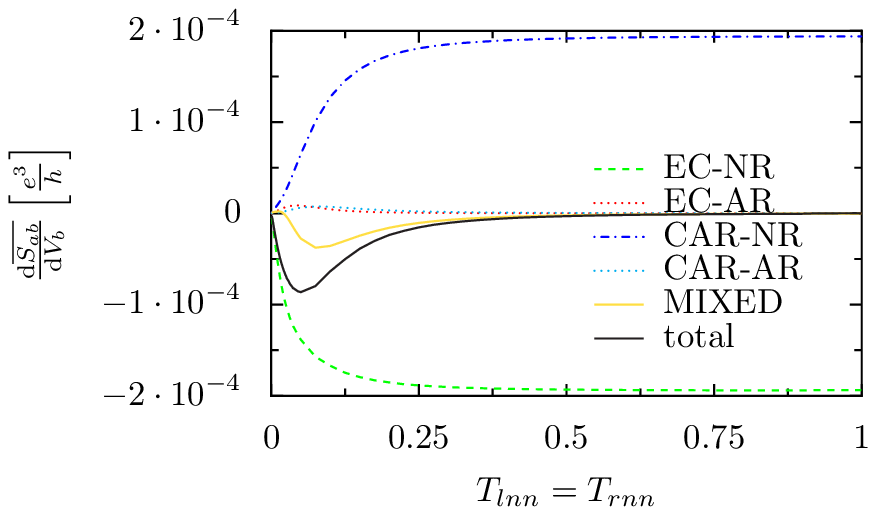}
\caption{\label{paperNSN:fig:SE-10} Averaged differential current cross-correlations in a) the
symmetrical bias situation $V_a=V_b\ll\Delta/|e|$ and b) 
the asymmetrical bias case $V_a=0$,
$V_b\ll\Delta/|e|$ for a superconducting electrode
shorter than the coherence length ($R=0.25\xi$) as a function of the
transparencies of the additional
barriers $T_{lnn}=T_{rnn}$. The barriers next to the superconductor are in the
tunnel regime ($T_{lns}=T_{rsn}=0.01$).}
\end{figure}

Let us turn back to independent averaging and have a look at the current  cross-correlations shown in figure~\ref{paperNSN:fig:SE-10}. In
the symmetrical bias case, the 
additional barriers do not lead to an enhancement of the signal. The noise is
dominated by the CAR-NR component,
and, as we have seen above, CAR is not influenced by reflectionless tunneling.
The EC-AR component is amplified by 
the additional barriers, because the AR amplitude describing a local process is amplified. This leads to a small shoulder in the total
cross-correlations. But since we are in the tunnel
regime and the leading order of CAR-NR is $T^2$ while the leading order of EC-AR
is $T^4$, the influence of the 
EC-AR-component is to small to lead to a global maximum. 

In the asymmetrical bias case $V_a=0,V_b\ll\Delta/|e|$, on the other hand,  the additional barriers enhance the
signal. But the cross-correlations are
dominated by EC-NR and are therefore negative.
We conclude that in a phase-averaged system, additional barriers only enhance the AR-component,
a local processes. It cannot help to amplify nonlocal signals. Thus, positive cross-correlations, when due to nonlocal
processes cannot be enhanced with additional barriers if an average over the length has to be done.
 
\section{The Ballistic System\label{sec:NSN:Ballistic}}
In a last step, let us examine if the additional barriers can enhance the current cross-correlations in a ballistic one-channel device, where no integrals over the length of the normal conducting channels need to be done. We nevertheless maintain he average over the central superconducting electrode. One can indeed imagine that the normal conducing channels are formed by single-wall nano-tubes, where transport is ballistic, while the superconducting central electrode is made from more disordered aluminum. 
\subsection{Resonances in a Fabry-Perot Interferometer\label{Ballistic:Sec:Intro}}
Assuming ballistic transport in the normal conducting parts, the NNSNN-system consists in principle of two Fabry-Perot interferometers, one at the left- and one at the right-hand side of the superconductor. For the final result the entire system has to be taken into account. But thinking of the two Fabry-Perot interferometers independently already gives a fair idea of where to expect resonance peaks, and what influence they have on the different components of the conductance and of the current cross-correlations.

In an optical Fabry-Perot interferometer consisting of two parallel mirrors, the transmitted intensity results from the interference of light which has undergone multiple reflections between the two mirrors. The transmission of a Fabry-Perot interferometer is a function of the  phase difference between a beam which has been $n$-times and a beam which has been $(n+1)$-times reflected at both interfaces.
The same is true for a Fabry-Perot interferometer for electrons. An electron going once back and forth between the two barriers at the left-hand side of the NNSNN-system acquires a phase $\phi^e=2q^+L_l$, a hole acquires a phase $\phi^h=2q^-L_l$, where $q^+$ is the wavevector of the electrons and $q^-$ the wavevector of the holes. For low voltages, where $q^+\approx q^-$, we expect electron and hole resonances to happen for the same resonator lengths. If the voltage is increased, the wavevector $q^+=\sqrt{2m\hbar^2}\sqrt{\mu-|e|V}$ decreases and the wave vector $q^-=\sqrt{2m\hbar^2}\sqrt{\mu+|e|V}$ increases, shifting the resonance for electrons to higher and the resonance for holes to lower values of $L_l$. The width of the resonances of a Fabry-Perot interferometer decreases as the reflectivity of the interfaces increases. 

\subsubsection{Symmetrically applied bias}
\begin{figure}
a)\includegraphics[width=0.45\columnwidth]{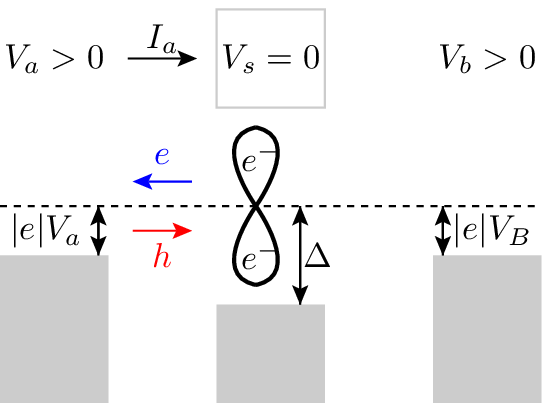}
b)\includegraphics[width=0.45\columnwidth]{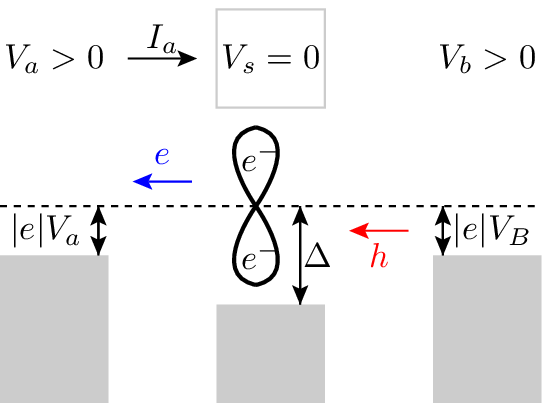}\\
c)\includegraphics[width=0.45\columnwidth]{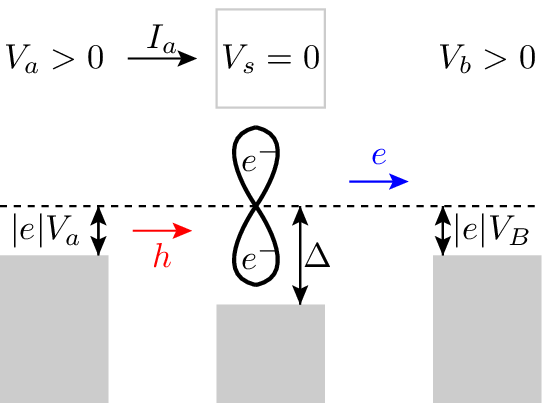} 
d)\includegraphics[width=0.45\columnwidth]{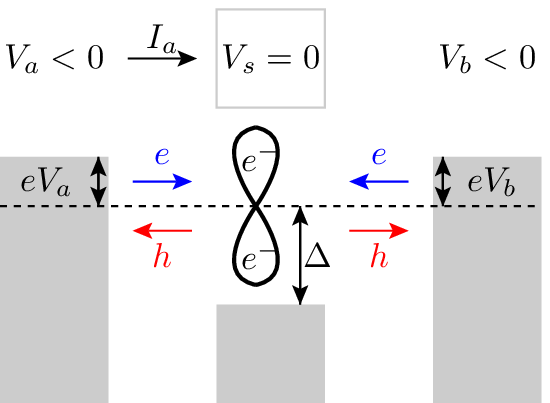}
\caption{\label{FP:fig:CondVaVb}Schematics of the contributions to the conductance in the symmetrically biased case. The filled states are depicted in the three reservoirs, together with the electron and hole currents between the barriers. a) AR contribution for positive bias, b) and c) The two possible CAR contributions for positive bias, d) Summary of the processes for negative bias. If negative voltages are applied, the roles of electrons and holes are interchanged, but the resulting conductance stays unchanged.}
\end{figure}
For a symmetrically applied bias, the differential conductance consists of an AR and a CAR component. 
If a positive bias is applied to the normal electrode $N_a$, the current $I_a$ contains an AR contribution, where a hole enters and an electron leaves the superconductor (see figure~\ref{FP:fig:CondVaVb}a). At low energies, the AR contribution will have a maximum for values of $L_l$ where electrons and holes are both in resonance. At higher energies, the resonances for electrons and for holes occur at different values of $L_l$. If the barriers are in the tunnel regime, the resonance peaks are sharp. The resonance peaks for electrons and holes stop overlapping and the AR contribution disappears. If the interface transparency is increased, the resonance peaks broaden. Even if the electron and the hole wavevectors are significantly different, there is an overlap between corresponding resonances leading to a double peak in the AR contribution.    
AR is a local process, and we expect the AR contribution to the current $I_a$ to be mainly independent of the length $L_r$.

There are two possibilities for CAR process if the electrodes $N_a$ and $N_b$
are symmetrically and positively biased: Either a hole enters the superconductor at
the left-hand side and an electron leaves the superconductor at the right-hand
side (see figure~\ref{FP:fig:CondVaVb}b); or the process occurs the other way
around (see figure~\ref{FP:fig:CondVaVb}c), the hole enters at the right-hand
side and the electron leaves at the left-hand side. In a CAR process, both sides
are involved and the position of the resonances will be a function of both $L_l$
and $L_r$. At low energies, the resonances for the two CAR processes will be
superimposed and occur at length values $L_l=L_r\!\!\mod\pi/k_F$. In a CAR
process, the incoming hole and the outgoing electron are spatially separated. It
is therefore possible to reach the optimal resonance condition for electrons and
for holes even if the two kinds of carriers have different wavevectors. For
higher bias values, the conductance maximum due to the CAR process sketched in
figure~\ref{FP:fig:CondVaVb}b will be shifted to lower values of $L_l$ and
higher values of $L_r$. It will be the other way around for the conductance
maximum due to the second CAR process. Therefore, we expect the resonance peak
to split up into two peaks.

CAR and AR both have a positive contribution to the differential conductance. This can also be seen in figure~\ref{FP:fig:CondVaVb}: The current $I_a$ is defined as positive if positive charges enter the superconductor. On the left-hand side, all electrons leave the superconductor and all holes enter it. 
If a negative bias is applied instead of a positive one, electrons and holes will interchange their roles (see figure~\ref{FP:fig:CondVaVb}d). The resonance positions and the sign of the conductance, however, will not change because $q^+(V)=q^-(-V)$.      

\subsubsection{Asymmetrically applied bias $V_a=0$, $V_b>0$}
\begin{figure}
a)\includegraphics[width=0.45\columnwidth]{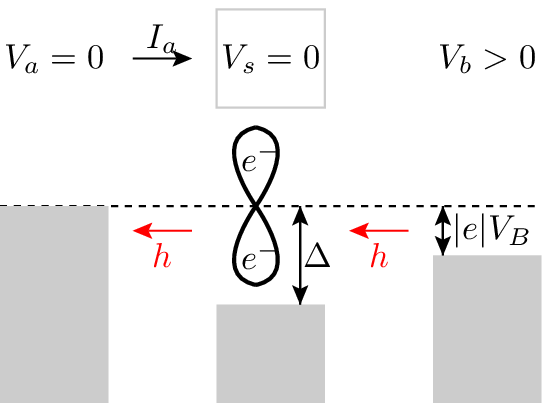} 
b)\includegraphics[width=0.45\columnwidth]{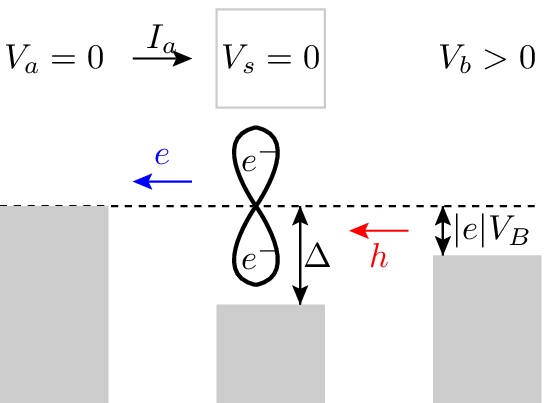}
\caption{\label{FP:fig:CondVb}Schematics of the contributions to the conductance in the asymmetrically biased configuration. The filled states are depicted in the three reservoirs, together with the current of electrons and holes between the barriers. a) EC contribution for positive bias, b) CAR contribution for positive bias.}
\end{figure}
Let us now focus on the bias configuration $V_a=0$, $V_b>0$ used to measure a nonlocal conductance.
The latter consists of an EC and a CAR contribution. EC processes involve the same type of charge carriers on both sides of the superconductor. Therefore, the resonances for EC will always occur for diagonal values $L_l=L_r\!\!\mod\pi/k_F$. At positive bias, EC is carried by holes (see figure~\ref{FP:fig:CondVb}a) and if the voltage is increased, the resonances in $L_l$ and $L_r$ are simultaneously shifted to lower values. EC leads to a negative differential conductance as the holes move out of the superconductor and the current is defined positive when entering the superconductor.

From the two kinds of possible CAR processes in the symmetrically biased case, only the second one contributes to the nonlocal conductance. At low energies the EC and the CAR peak will be superimposed. At higher energies, the CAR peak moves away from the diagonal. 
The application of a negative bias again interchanges the roles of electrons and holes, but it does not influence the resonance positions. The only difference between a positive and a negative applied bias is the global sign of the current cross-correlations.

If one wishes to use a Fabry-Perot interferometer as a filter for the different processes contributing to the conductance and to the current cross-correlations, we expect that it will be essential to operate the system at energies where the wavevectors of electrons and holes are significantly different.
In the next section, numerical results will be analyzed, which will turn out to be in qualitative agreement with the scenario discussed in this section. We start with the tunnel regime, then investigate how the contrast diminishes with increasing interface transparency and eventually have a closer look at a system with intermediate interface transparencies.

\subsection{The Tunnel Regime}
Let us now discuss the calculated conductance and the current cross-correlations. In a first step, the barriers between the superconducting and the normal regions are chosen to be in the tunnel regime ($T_{lns}=T_{rsn}\approx0.01$) in order to achieve sharp resonances. The distances $L_l$ and $L_r$ are of the order of ten times the coherence length $\xi$. This distance allows to observe phase differences $\Delta\phi=|\phi^e-\phi^h|$ of the order of $2\pi$ between electrons and holes already at energies much smaller than the superconducting gap.

\begin{figure}
\begin{minipage}{0.48\columnwidth}
a)\includegraphics[width=0.99\textwidth]{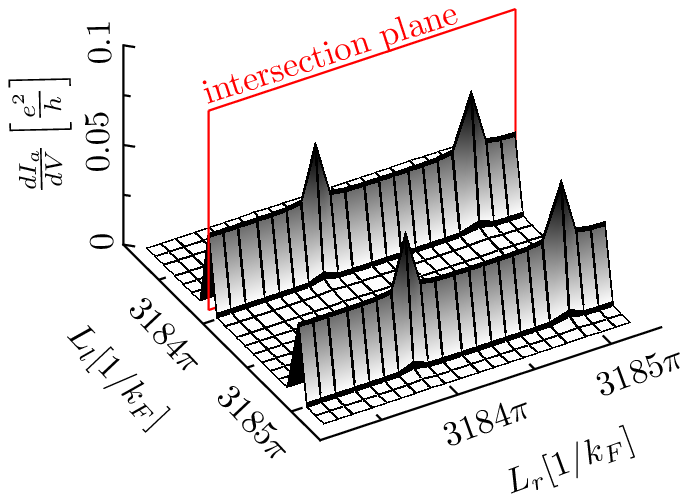}\\
b)\includegraphics[width=0.95\textwidth]{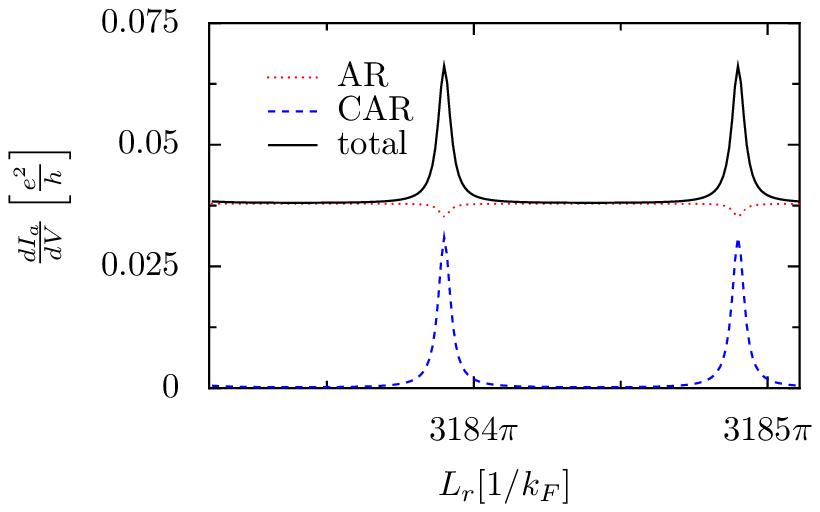}
\end{minipage}
\begin{minipage}{0.48\columnwidth}
c)\includegraphics[width=0.99\textwidth]{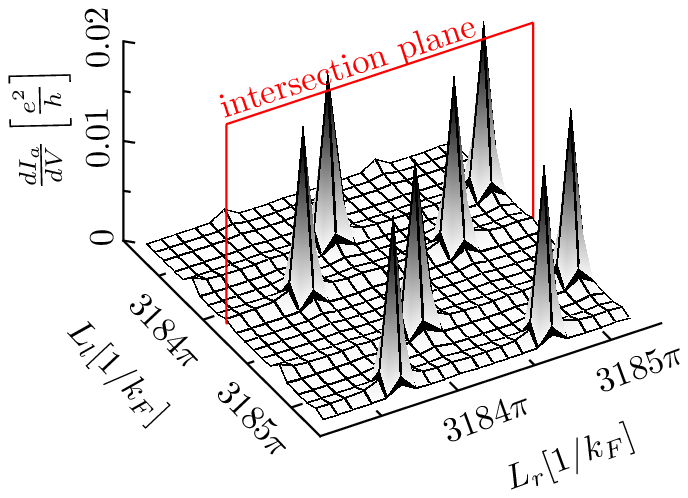}\\
d)\includegraphics[width=0.95\textwidth]{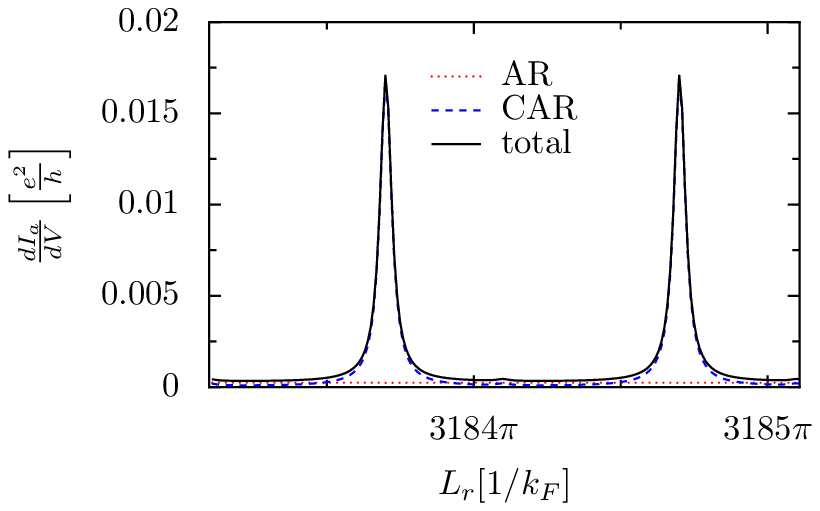}
\end{minipage}
\caption{\label{FP:fig:DiffCondVaVb}Differential conductance for symmetrically applied bias $V=V_a=V_b$ in the tunneling limit $T_{lns}=T_{rsn}=0.01$, $T_{lnn}=T_{rnn}=0.26$,
a) as a function of $L_r$ and $L_l$ at low voltage $|e|V=1\cdot10^{-4}\Delta$,  
b) cut through a) along one of the "mountain ranges" (at ${L_l=3183.9\pi/k_F}$) at higher resolution,
c) as a function of $L_r$ and $L_l$ at higher voltage $|e|V=6.25\cdot10^{-2}\Delta$,
d) cut through c) at $L_l=3184.1\pi/k_F$ at higher resolution.}
\end{figure}

\begin{figure}
\begin{minipage}{0.48\columnwidth}
a)\includegraphics[width=0.99\textwidth]{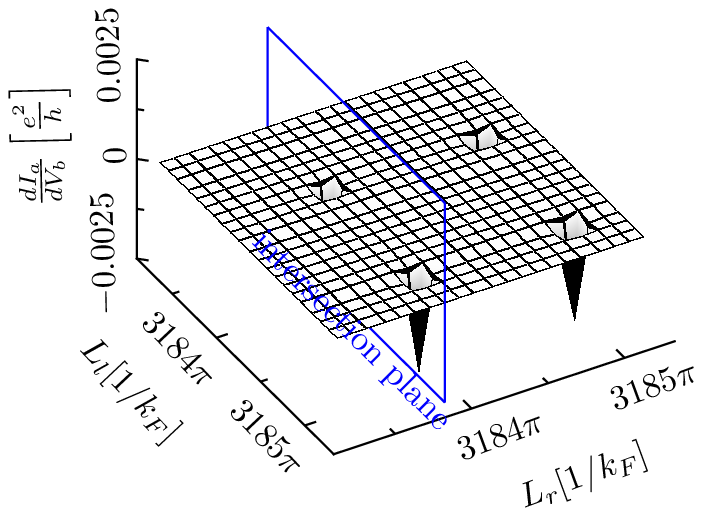}\\
b)\includegraphics[width=0.95\textwidth]{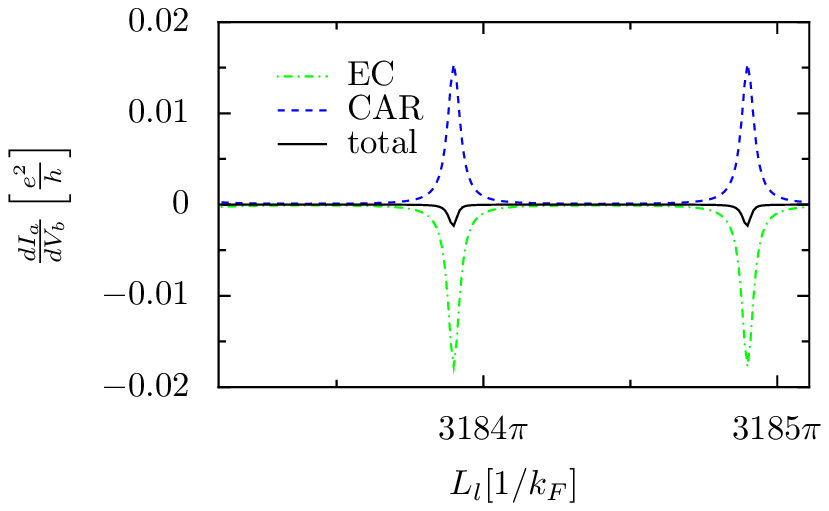}
\end{minipage}
\begin{minipage}{0.48\columnwidth}
c)\includegraphics[width=0.99\textwidth]{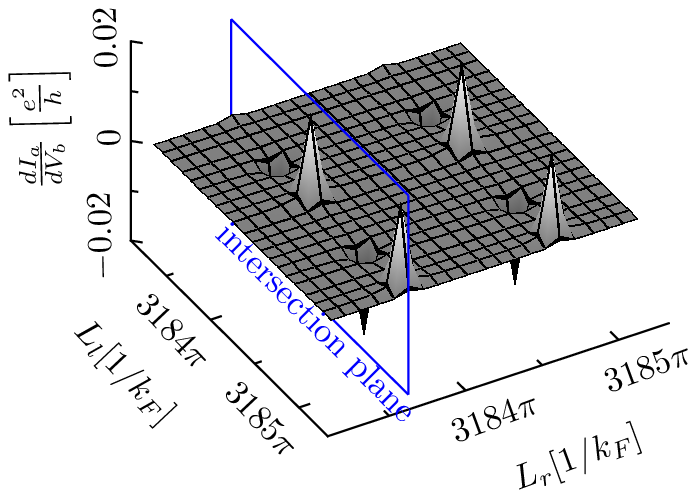}\\
d)\includegraphics[width=0.95\textwidth]{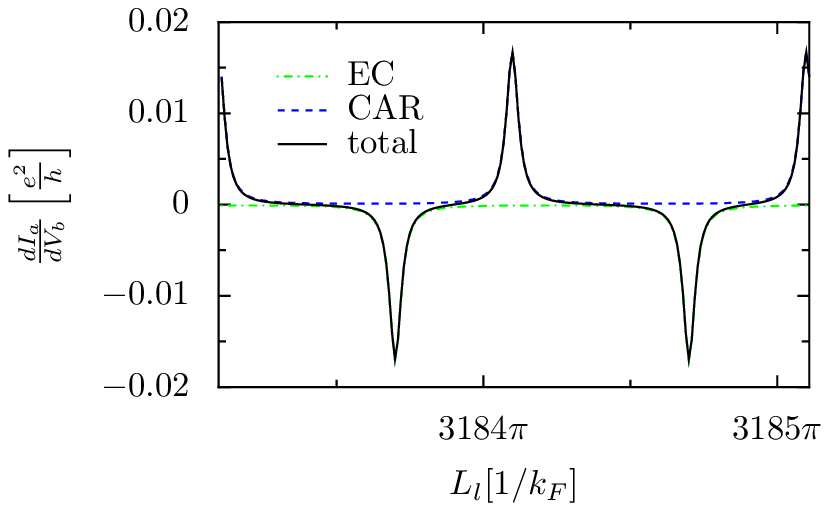}
\end{minipage}
\caption{\label{FP:fig:DiffCondVb}Differential nonlocal conductance ($V_a=0$) in the tunneling limit $T_{lns}=T_{rsn}=0.01$, $T_{lnn}=T_{rnn}=0.26$,
a) as a function of $L_r$ and $L_l$ at low voltage $|e|V_b=1\cdot10^{-4}\Delta$,  
b) cut through a) at ${L_r=3183.9\pi/k_F}$ at higher resolution,
c) as a function of $L_r$ and $L_l$ at higher voltage $|e|V_b=6.25\cdot10^{-2}\Delta$,
d) cut through c) at $L_r=3183.6\pi/k_F$ at higher resolution.}
\end{figure}

\begin{figure}
\begin{minipage}{0.48\columnwidth}
a)\includegraphics[width=0.99\textwidth]{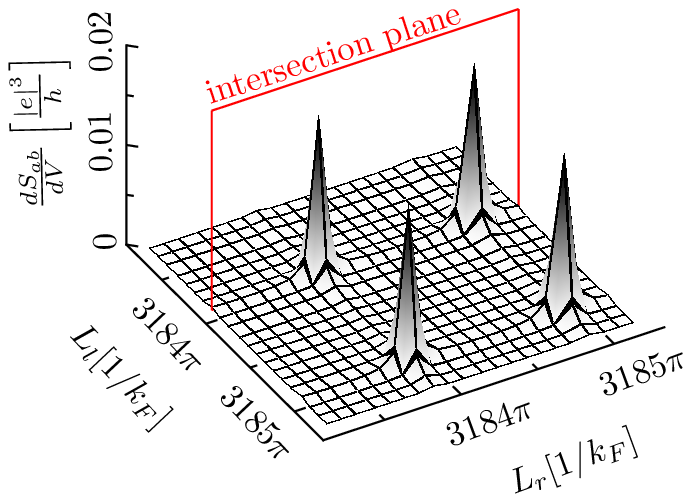}
b)\includegraphics[width=0.95\textwidth]{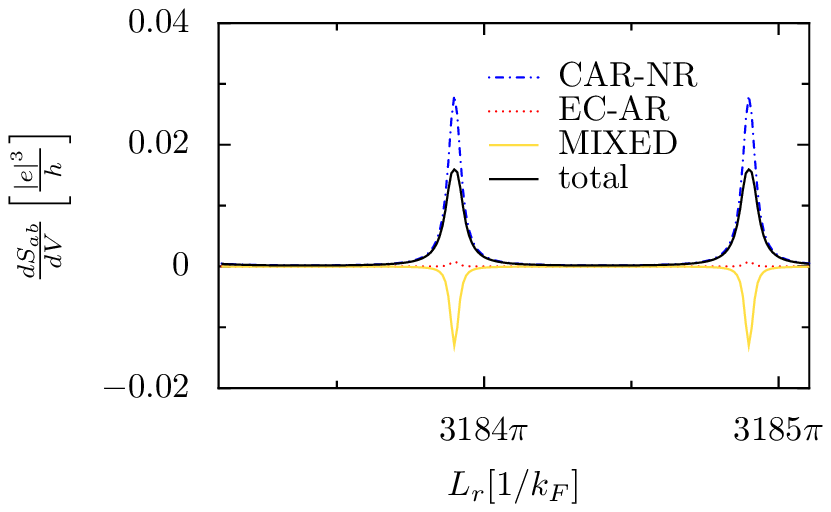}
\end{minipage}
\begin{minipage}{0.48\columnwidth}
c)\includegraphics[width=0.99\textwidth]{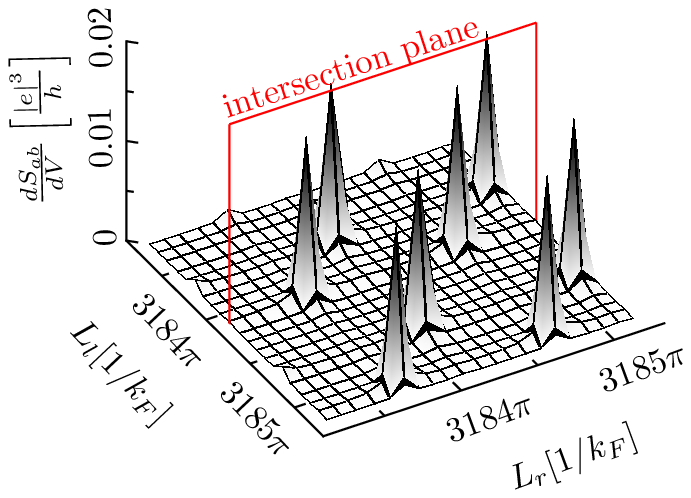}
d)\includegraphics[width=0.95\textwidth]{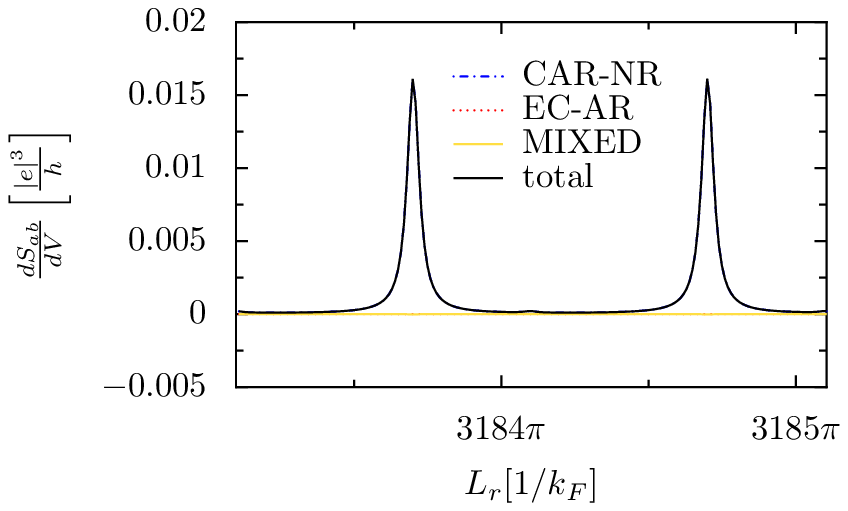}
\end{minipage}
\caption{\label{FP:fig:NoiseVaVb}Differential current cross-correlations for $V_a=V_b$ in the tunneling limit $T_{lns}=T_{rsn}=0.01$, $T_{lnn}=T_{rnn}=0.26$,
a) as a function of $L_r$ and $L_l$ at low voltage $|e|V=1\cdot10^{-4}\Delta$,  
b) cut through a) at ${L_l=3183.9\pi/k_F}$ at higher resolution,
c) as a function of $L_r$ and $L_l$ at higher voltage $|e|V=6.25\cdot10^{-2}\Delta$,
d) cut through c) at $L_l=3184.1\pi/k_F$ at higher resolution. }
\end{figure}

\begin{figure}
\begin{minipage}{0.48\columnwidth}
a)\includegraphics[width=0.95\textwidth]{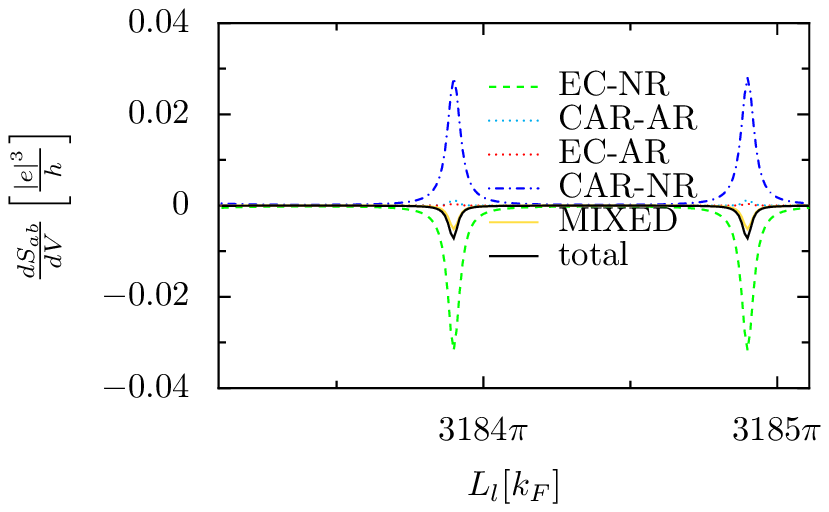}
\end{minipage}
\begin{minipage}{0.48\columnwidth}
b)\includegraphics[width=0.95\textwidth]{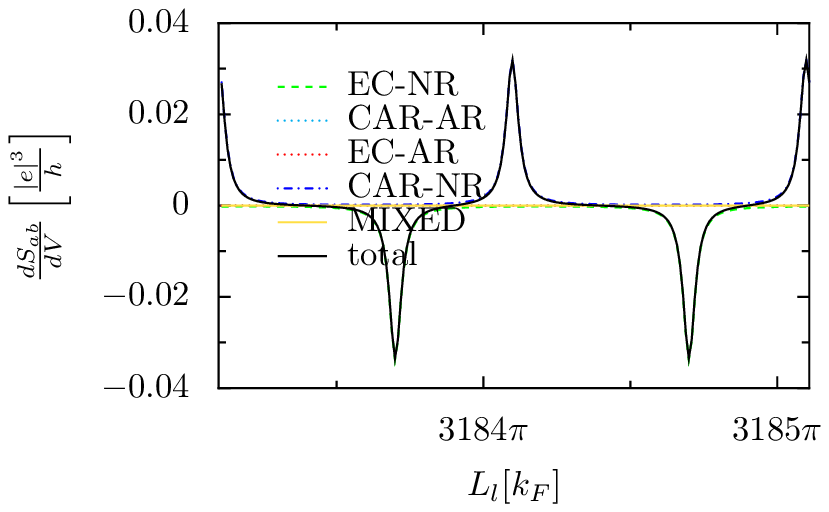}
\end{minipage}
\caption{\label{FP:fig:NoiseVb}Differential current cross-correlations for $V_a=0$ in the tunneling limit $T_{lns}=T_{rsn}=0.01$, $T_{lnn}=T_{rnn}=0.26$. The three-dimensional plots for the current cross-correlations are quasi-identical to the ones for the nonlocal conductance in figure~\ref{FP:fig:DiffCondVb}, therefore only the cuts are shown.  
a) ${L_r=3183.9\pi/k_F}$, $|e|V=1\cdot10^{-4}\Delta$,  
b) $L_r=3183.6\pi/k_F$, $|e|V_b=6.25\cdot10^{-2}\Delta$}
\end{figure}

As for the averaged system, we will 
study four different quantities: the conductance in the symmetrically biased case, the nonlocal conductance, where the normal electrode in which the current is calculated is at the same potential as the superconductor, and the current cross-correlations in the same two bias configurations. The results are presented in the figures~\ref{FP:fig:DiffCondVaVb}, \ref{FP:fig:DiffCondVb}, \ref{FP:fig:NoiseVaVb} and \ref{FP:fig:NoiseVb}. The panels a) and c) always show a three-dimensional plot of the conductance (respectively the current cross-correlations) as a function of the distances between the barriers $L_l$ and $L_r$. The three-dimensional plot in panel a) always depicts the situation in the limit of zero energy. The one in panel c) refers to higher energies. The three-dimensional plots are completed by cuts through the resonances. These cuts show the different contributions in different colors and have a higher resolution.

\subsubsection{Conductance for a symmetrically biased system}
Figure~\ref{FP:fig:DiffCondVaVb}a shows the differential conductance $\left.dI_a/dV\right|_{V=V_a=V_b>0}$ in the symmetrically biased case $V_a=V_b$ as a function of the distances $L_l$ and $L_r$ in the limit $V\to0$. There are resonance "mountain ranges" for two values of $L_l$ and there are four peaks which stand out of the mountain ranges. 
The mountain ranges correspond to a $L_r$-independent process and will therefore be due to AR. This is confirmed by figure~\ref{FP:fig:DiffCondVaVb}b showing in a cut along one of the mountain ranges the AR and the CAR component separately. 
Figure~\ref{FP:fig:DiffCondVaVb}b also shows that the mountain peaks are due to CAR.
The relative distance of two subsequent resonance peaks in units of $k_F$ is $\pi$ in $L_r$ and in $L_l$ direction. But the absolute resonance positions are not given by integer multiples of $\pi$ as it would be the case in a simple optical Fabry-Perot interferometer with two identical mirrors. 
Figure~\ref{FP:fig:DiffCondVaVb}c shows the differential conductance at higher voltage (but still much smaller than the gap). As expected from section~\ref{Ballistic:Sec:Intro}, the AR mountain ranges disappear because the electron and hole resonances do not overlap anymore. Every CAR peak splits up into two peaks corresponding to the two CAR contributions.

The observation of the disappearance of AR and the splitting of the CAR peaks is already interesting in itself. 
As the peaks are uniquely due to CAR processes, the desired filtering of the different processes contributing to the conductance is achieved. A NNSNN-system operated with symmetrical bias at energies where AR is suppressed and tuned to a CAR resonance peak could be a good source of entangled electrons.

\subsubsection{Non-local conductance for $V_a=0$, $V_b>0$}
 Figure~\ref{FP:fig:DiffCondVb} depicts the calculated differential nonlocal conductance for the biases $V_a=0$, $V_b>0$. Panel~a) shows the nonlocal differential conductance $\left.dI_a/dV_b\right|_{V_a=0}$ as a function of the distances $L_l$ and $L_r$ at very low voltage $V_b$. There are two resonance peaks on the diagonal, where $L_l=L_r$, and two additional resonance peaks where $L_l=L_r\pm\pi/k_F$. From the considerations in the last section, we expect the EC and the CAR resonances to appear at the same positions. As the peaks are negative, the EC contribution is apparently stronger than the CAR contribution. Figure~\ref{FP:fig:DiffCondVb}b showing a cut through figure~\ref{FP:fig:DiffCondVb}a at constant $L_l$, where the total current is decomposed into its EC and CAR component, confirms that the peaks in the three-dimensional plot are indeed a superposition of EC and CAR.
 
When $V_b$ is increased, the EC and the CAR resonance move apart (see figure~\ref{FP:fig:DiffCondVb}c). 
As expected, the EC peaks stay on the diagonals and at $L_l=L_r\pm\pi/k_F$, but their positions are shifted to lower length values. The positive conductance peaks due to CAR
appear at the same $L_r$ value as the EC resonance peaks, as both processes are carried by holes on the right hand side. On the left hand side CAR is carried by electrons whose wave vector $q^+$ decreases with increasing voltage $V_b$. Consequently, the $L_l$ values of the peak position increase with increasing voltage.
Again, the desired filtering of different processes is obtained. This voltage configuration is interesting as the conductance switches between positive and negative values if $L_r$ is changed.
\subsubsection{Differential current cross-correlations}
The differential current cross-correlations $\left.dS_{ab}/dV\right|_{V_a=V_b=V}$ depicted in figure~\ref{FP:fig:NoiseVaVb} and $\left.dS_{ab}/dV_b\right|_{V_a=0}$ depicted in figure~\ref{FP:fig:NoiseVb} behave very similarly to the corresponding differential conductance if one compares the CAR-NR component of the current cross-correlations with the CAR component of the conductance and the EC-NR component with the EC component. The only striking difference is the absence of the "mountain range" in the noise at low symmetrical bias. The similarity between current cross-correlations and conductance has already been discussed at the end of section~\ref{paperNNSNN:sec:comp} in connection to the work of Bignon \textit{et\,al.}~\cite{bignon}: As the barriers are opaque, the most likely process to occur is normal reflection. Therefore, the elements of the scattering matrix corresponding to normal reflection are close to one and EC-NR is similar to EC and CAR-NR is similar to CAR.
Only the AR component of the conductance has no counterpart in the current cross-correlations, because
all elements of the current cross-correlations are composed of a local and a nonlocal process. An AR-NR component does not exist and the AR mountain range is absent in the noise.
\subsection{Increasing the Interface Transparency}
\begin{figure}
\includegraphics[width=0.48\columnwidth]{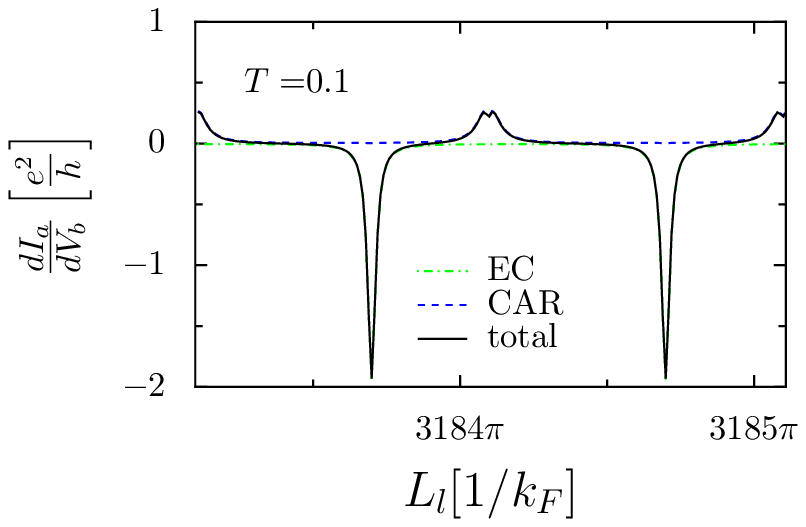}
\includegraphics[width=0.48\columnwidth]{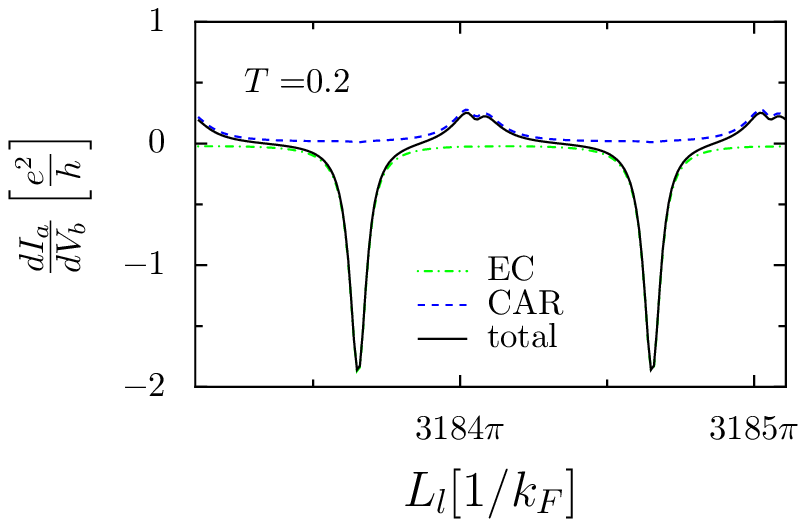}\\
\includegraphics[width=0.48\columnwidth]{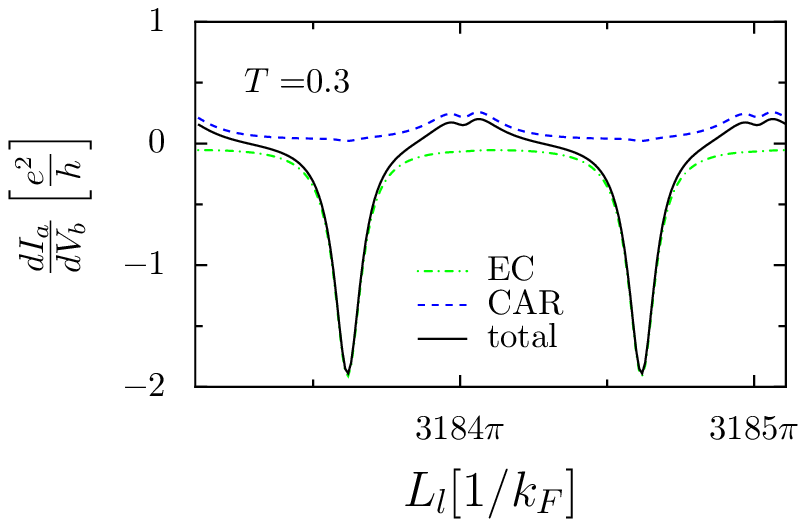}
\includegraphics[width=0.48\columnwidth]{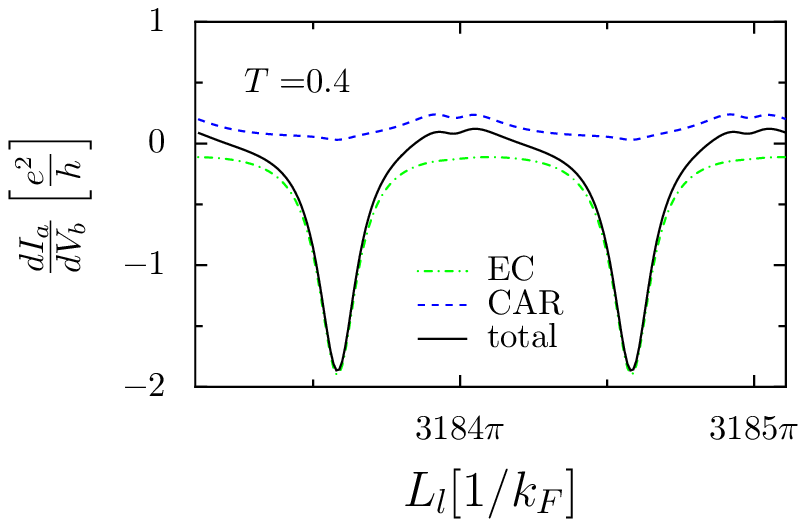}\\
\includegraphics[width=0.48\columnwidth]{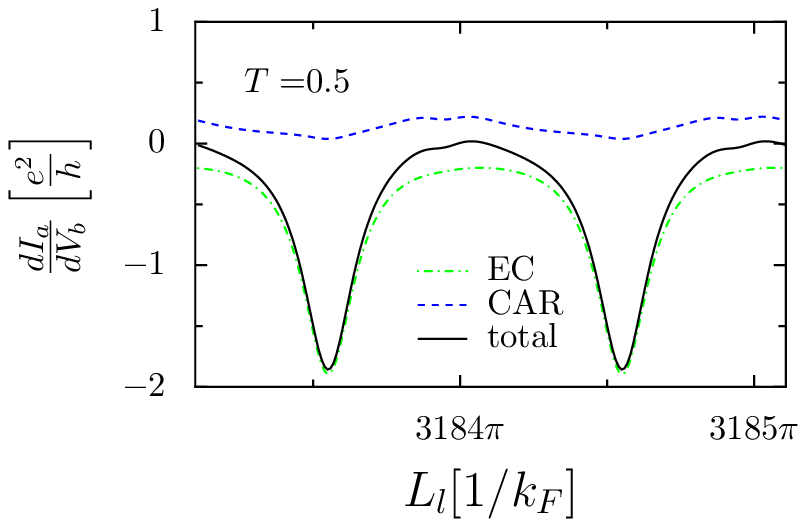}
\includegraphics[width=0.48\columnwidth]{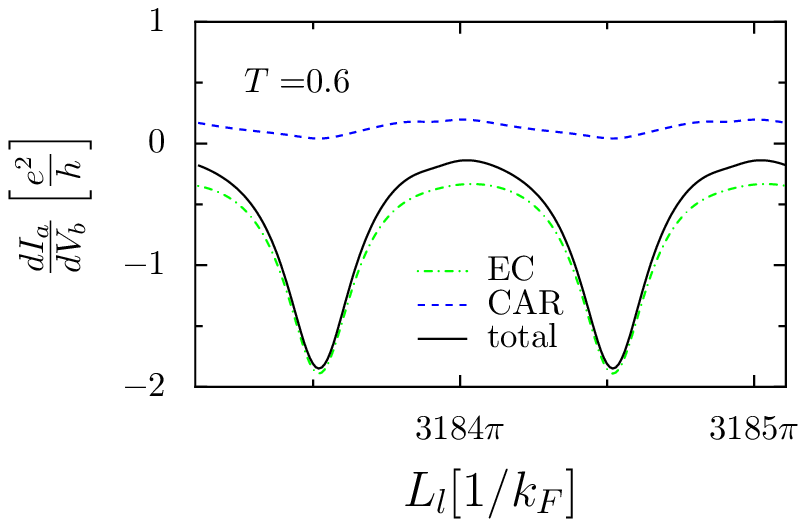}
\caption{\label{NSN:fig:LRScanR0.25}Non-local conductance at $|e|V_b=6.25\cdot10^{-2}\Delta$ and $V_a=0$ for $T_{lnn}=T_{lns}=T_{rsn}=T_{rnn}=T$ as indicated in the figure, for a very short superconducting electrode $R=0.25\xi$, at constant $L_r$ value, chosen to be in resonance so that the conductance in function of $L_l$ features CAR and EC peaks. The CAR peak splits up in two peaks. For $T=0.5$ and larger, the CAR peak has completely disappeared.}
\end{figure}
\begin{figure}
\includegraphics[width=0.48\columnwidth]{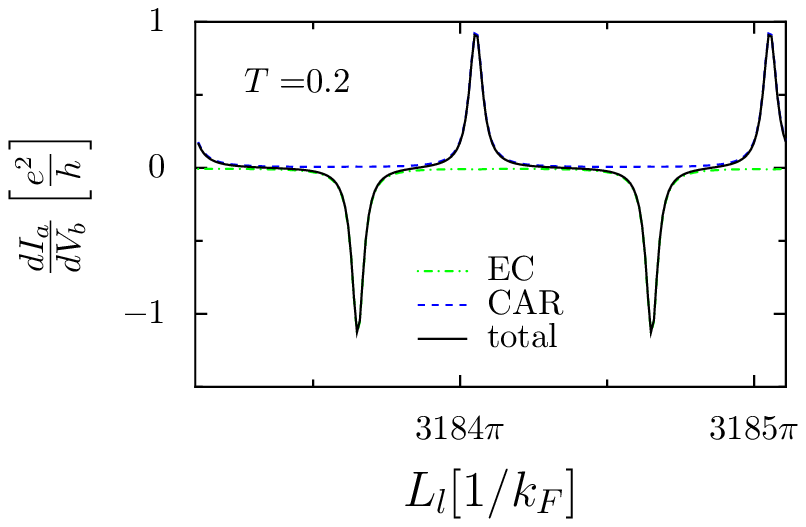}
\includegraphics[width=0.48\columnwidth]{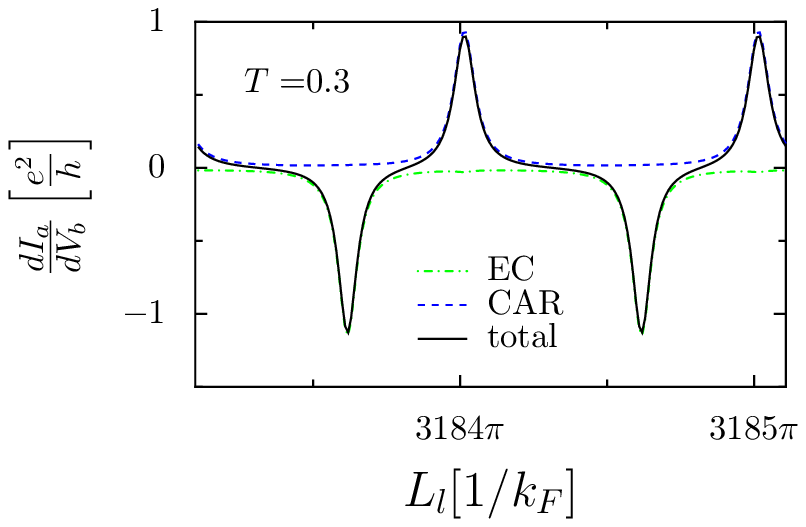}\\
\includegraphics[width=0.48\columnwidth]{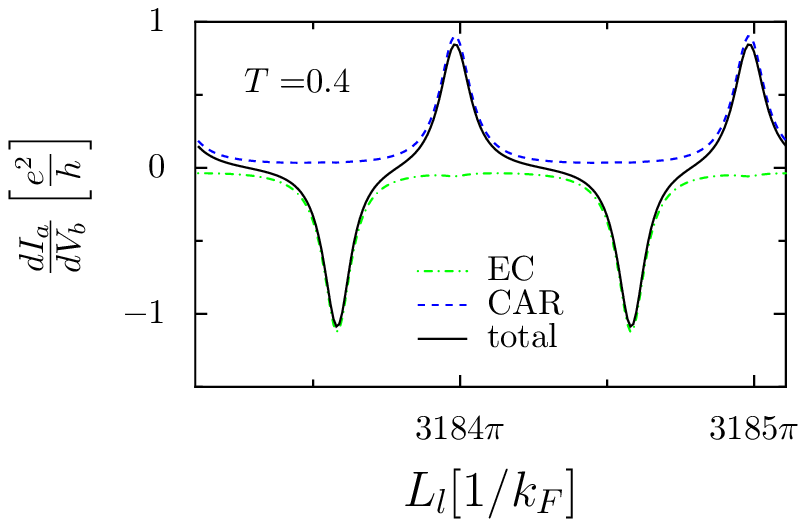}
\includegraphics[width=0.48\columnwidth]{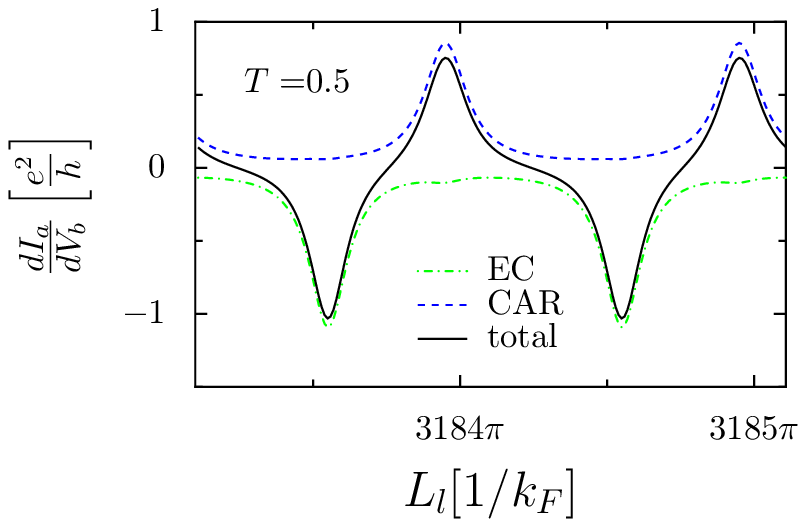}\\
\includegraphics[width=0.48\columnwidth]{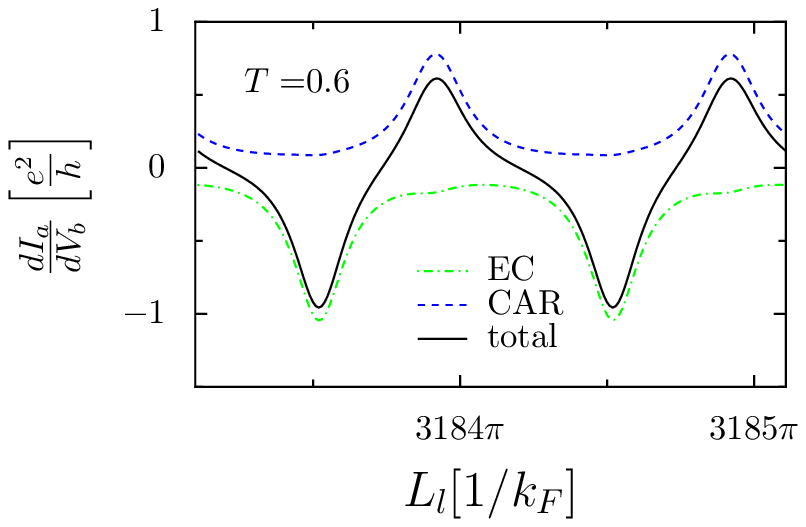}
\includegraphics[width=0.48\columnwidth]{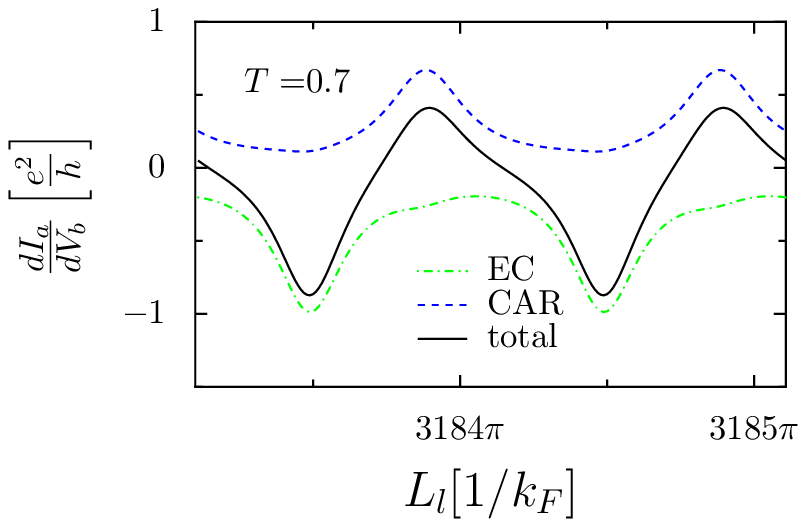}
\caption{\label{NSN:fig:LRScanR1}Non-local conductance at $eV_b=6.25\cdot10^{-2}\Delta$ and $V_a=0$ for $T_{lnn}=T_{lns}=T_{rsn}=T_{rnn}=T$ as indicated in the figure, for a slightly longer superconducting electrode $R=\xi$, at constant $L_r$ value, chosen to be in resonance so that the conductance in function of $L_l$ features CAR and EC peaks. The EC  and CAR peaks are sharper than in the case of the extremely short superconducting electrode depicted in figure~\ref{NSN:fig:LRScanR0.25} and there is no splitting. Even for barrier transparencies as high as $T=0.7$ the signature of the EC and the CAR contribution is clearly visible.}
\end{figure}
In the last section, we have studied the ballistic system in the tunnel regime where all resonance peaks are sharp. However, the systems in which electronic Fabry-Perot interferences have already been observed, for example~\cite{liang2001fabry,PhysRevLett.96.207003}, were in the opposite regime of high interface transparency. 
For increased barrier transparencies, one expects the resonance peaks to become broader and the separation of the different peaks to become less pronounced. This section discusses whether it is necessary to stay in the tunnel regime or if the filtering effect survives to higher interface transparencies.
 
 To investigate systematically what happens when the interface transparency is increased, the nonlocal conductance at $eV_b=6.25\cdot10^{-2}\Delta$ and $V_a=0$ was calculated for different barrier transparencies, at constant $L_r$ value, which was chosen to be in resonance so that the conductance in function of $L_l$ would feature CAR  and EC peaks. Figure~\ref{NSN:fig:LRScanR0.25} shows this conductance for a short superconducting electrode $R=0.25\xi$ and for different values of $T_{lnn}=T_{lns}=T_{rsn}=T_{rnn}=T$. The CAR peak shows an interesting feature: It splits up into two peaks, while the EC peak does not. 
One possible explanation is  the interplay between the resonance in the NNS double barrier and the resonance in the central superconductor, both involving Andreev reflections. The first one implies De Gennes-Saint-James bound states \cite{saint-james}, broadened by the presence of infinite reservoirs. The second are the Tomasch-Rowell-McMillan \cite{tomasch} resonances, due to a scattering between quasiparticle states with different wavevectors $k^+,k^-$ in S.  In a NS bilayer, these two processes can altogether lead to a doubling of the resonance peaks \cite{entin-wohlmann}. In the present case, such a doubling is observed in the CAR channel, but is absent in the EC channel where all wavevectors are approximately equal. Moreover, higher transmissions favour the EC channel, and the CAR peak gets weakened altogether.
 
The amplitude of the quasiparticle wavefunction decreases exponentially with the superconducting length. Therefore, one can hope to decrease the coupling between the two resonances by elongating the superconducting electrode. This actually works, as the plots in figure~\ref{NSN:fig:LRScanR1} where $R=\xi$ show: The EC  and CAR peaks are sharper than in the case of the extremely short superconducting electrode and there is no splitting. Even for barrier transparencies as high as $T=0.7$, the signature of the EC and the CAR contribution is clearly visible.
Against the general trend, we have here a situation where a longer distance in the superconductor leads to a larger signal. 

In conclusion, it is not necessary to operate the NNSNN device in the tunnel regime in order to filter the different processes. Surprisingly, CAR resonances are better resolved if the distance in the superconducting electrode is not too short.
  
Let us now study a system for intermediate interface transparencies $T=T_{lnn}=T_{lns}=T_{rsn}=T_{rnn}=0.5$ and $R=\xi$, which is clearly outside the tunnel regime but has still a good resolution between the different processes, in more detail. We proceed in the same way as in the tunnel regime and analyze the conductance and the current cross-correlations in two different voltage configurations.
\subsubsection{Conductance for a symmetrically biased system}
\begin{figure}
\begin{minipage}{0.48\columnwidth}
a)\includegraphics[width=0.99\textwidth]{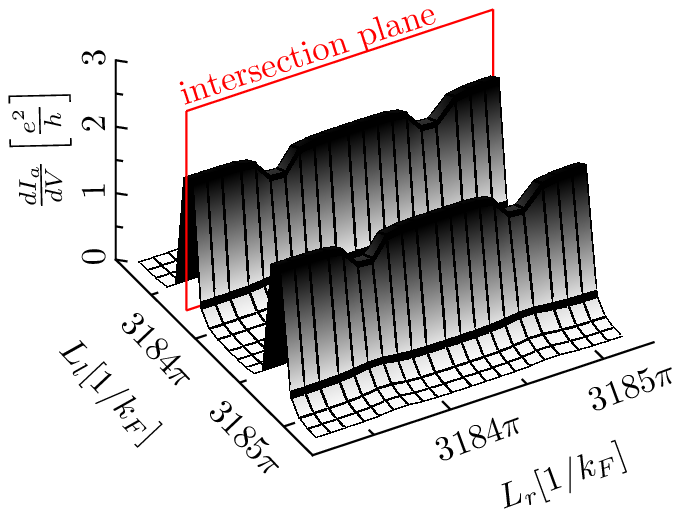}\\
b)\includegraphics[width=0.95\textwidth]{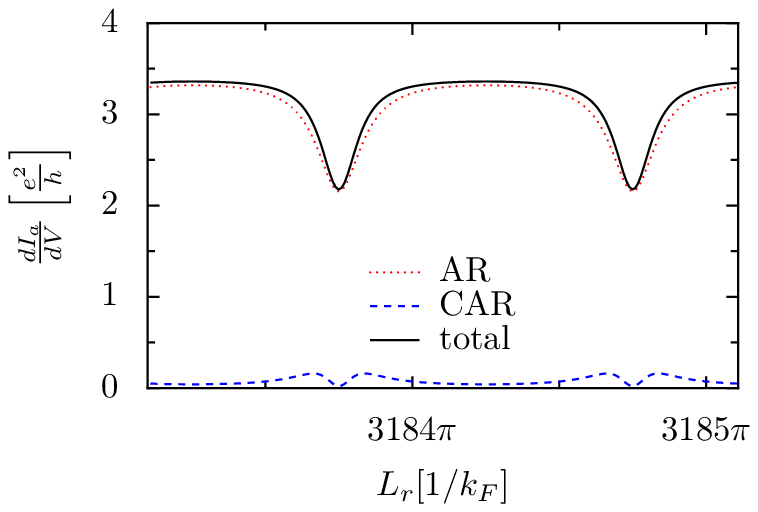}
\end{minipage}
\begin{minipage}{0.48\columnwidth}
c)\includegraphics[width=0.99\textwidth]{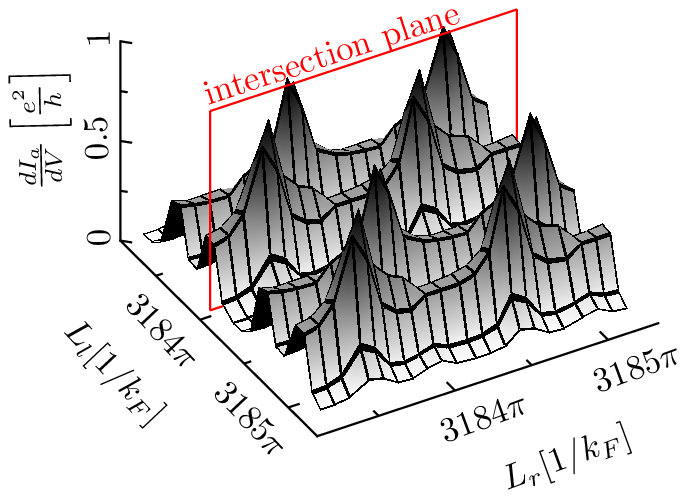}\\
d)\includegraphics[width=0.95\textwidth]{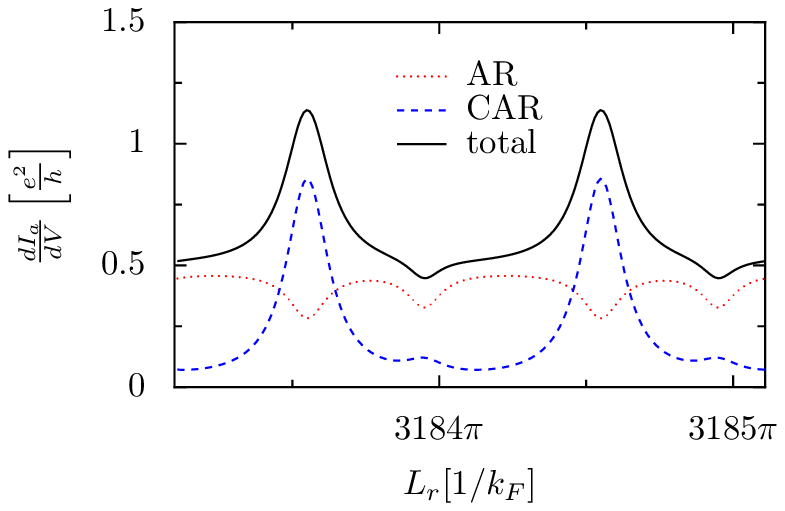}
\end{minipage}
\caption{\label{FP:fig:DiffCondVaVbR1T0p5}Differential conductance for symmetrically applied bias $V=V_a=V_b$ for intermediate interface transparency $T=T_{lnn}=T_{lns}=T_{rsn}=T_{rnn}=0.5$ and $R=\xi$, 
a) as a function of $L_r$ and $L_l$ at low voltage $|e|V=1\cdot10^{-4}\Delta$,
b) cut through a) along one of the mountain ranges (at ${L_l=3183.7\pi/k_F}$) at higher resolution,
c) as a function of $L_r$ and $L_l$ at higher voltage $|e|V=6.25\cdot10^{-2}\Delta$,
d) cut through c) at $L_l=3183.9\pi/k_F$ at higher resolution.
}
\end{figure}

Figure~\ref{FP:fig:DiffCondVaVbR1T0p5} shows the differential conductance for a symmetrically biased system. At low energies (figure~\ref{FP:fig:DiffCondVaVbR1T0p5}a) the predominant structures are again the AR "mountain ranges". But in contrast to the situation in the tunnel regime, there are no CAR peaks on top of the mountain ranges. Instead, at the positions we would have expected them, there are depressions. The cut along the mountain range (figure~\ref{FP:fig:DiffCondVaVbR1T0p5}b) shows that the AR component features a deep depression where $L_r=L_l$. This depression already exists in the tunnel regime (figure~\ref{FP:fig:DiffCondVaVb}b), but is much less pronounced there. The CAR component increases when $L_r$ approaches $L_l$ as if a peak was to form, but falls back to zero at $L_r=L_l$. 
We can understand these dips in the conductance from the competition of the different processes at the level of the scattering matrix.
At $L_r=L_l=3183.7\pi/k_F$, EC processes are in resonance (see section~\ref{paperNSN:sec:nonlocalhighT}). As the probabilities for the four possible processes have to sum up to one, a large EC scattering matrix element is only possible at the cost of the CAR and AR scattering matrix elements. In the tunnel regime, the most probable process is normal reflection. The EC scattering matrix element can be increased at the cost of the NR scattering matrix element which has no signature in the conductance. But here, the AR contribution is as high as $3.3e^2/h$, while the maximal possible value in a four channel system is $4e^2/h$.

If higher voltages are applied (see figure~\ref{FP:fig:DiffCondVaVbR1T0p5}c), each mountain range splits up into two less high mountain ranges. In the tunnel regime, the AR contribution disappears at higher voltage. Electrons and holes have different wavevectors and therefore different resonance lengths and if the resonances are narrow, electrons and holes are not in resonance at the same time. Here with the higher interface transparencies, the resonances are wider and if the electron is in resonance the hole part is still large enough to form a peak and vice versa. At higher applied voltage, the mountain ranges are surmounted by CAR peaks. The resonance of the EC elements of the scattering matrix occur at different values of $L_r$ and CAR and EC are not in direct competition. The AR contribution (figure~\ref{FP:fig:DiffCondVaVbR1T0p5}~d) features at higher voltage two depressions, where there was only one at lower voltage. In one case AR is decreased in favor of CAR and in the second case, on the diagonals where $L_r=L_l\mod\pi/k_F$, it is decreased, as in the low energy case, in favor of EC.

\begin{figure}
\begin{minipage}{0.48\columnwidth}
a)\includegraphics[width=0.99\textwidth]{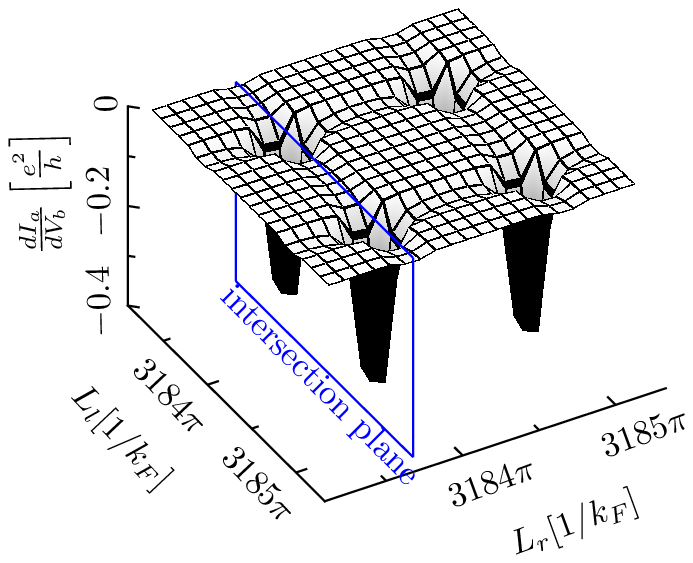}\\
b)\includegraphics[width=0.95\textwidth]{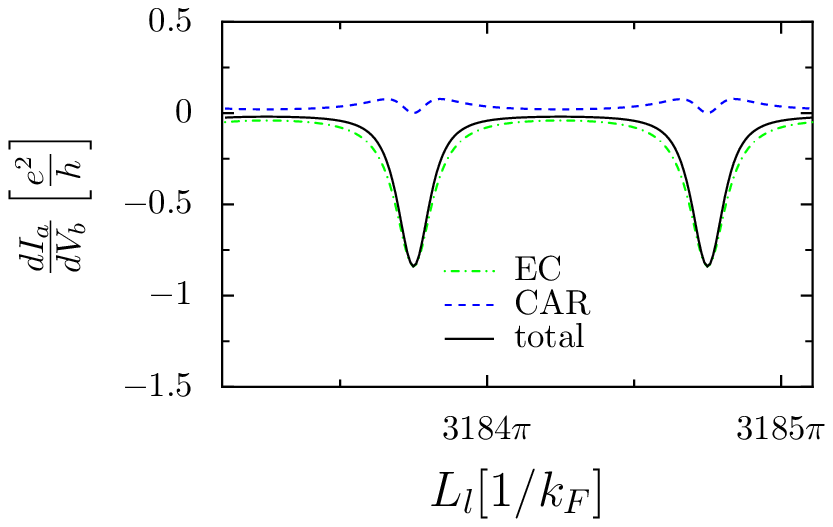}
\end{minipage}
\begin{minipage}{0.48\columnwidth}
c)\includegraphics[width=0.99\textwidth]{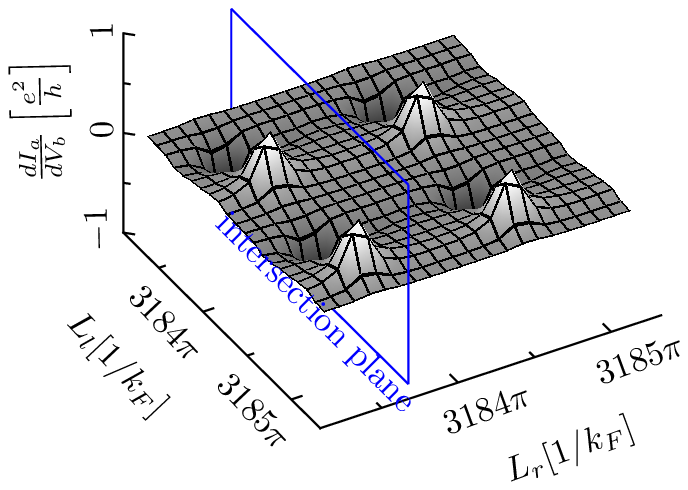}\\
d)\includegraphics[width=0.95\textwidth]{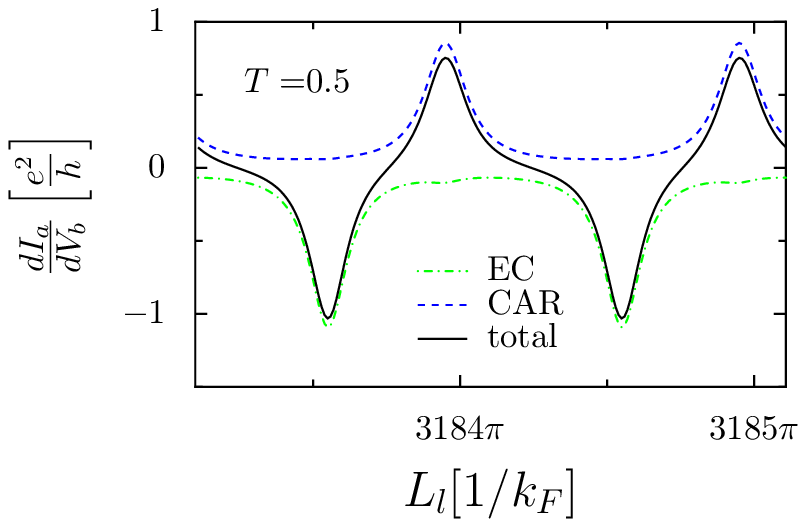}
\end{minipage}
\caption{\label{FP:fig:DiffCondVbR1T0p5}Differential nonlocal conductance $V_a=0$ for intermediate interface transparency $T=T_{lnn}=T_{lns}=T_{rsn}=T_{rnn}=0.5$ and $R=\xi$, 
a) as a function of $L_r$ and $L_l$ at low voltage $|e|V_b=1\cdot10^{-4}\Delta$,  
b) cut through a) at ${L_l=3183.7\pi/k_F}$ at higher resolution,
c) as a function of $L_r$ and $L_l$ at higher voltage $|e|V_b=6.25\cdot10^{-2}\Delta$,
d) cut through c) at $L_l=3183.9\pi/k_F$ at higher resolution.
}
\end{figure}
\begin{figure}
\begin{minipage}{0.48\columnwidth}
a)\includegraphics[width=0.99\textwidth]{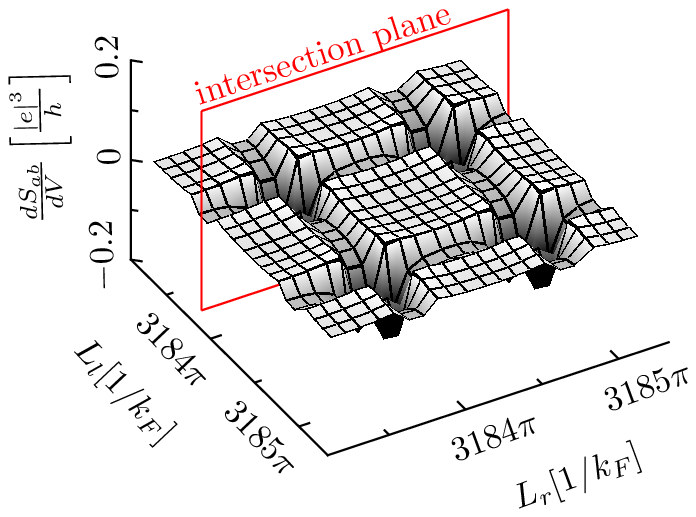}\\
b)\includegraphics[width=0.95\textwidth]{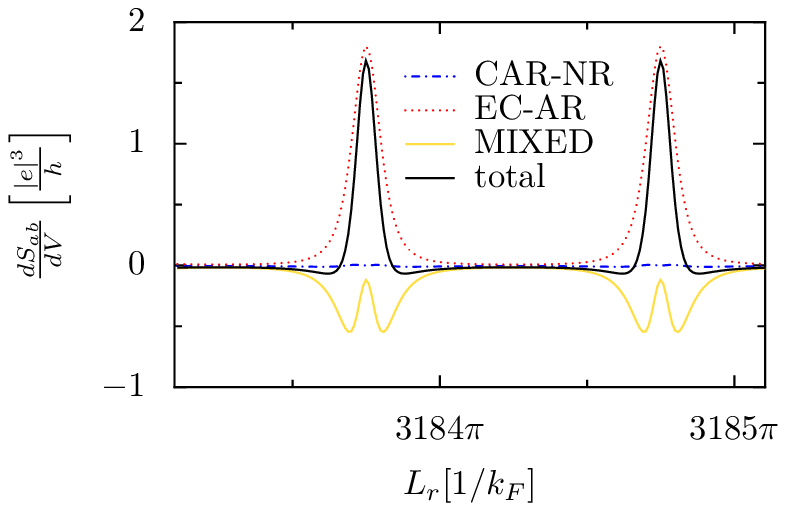}
\end{minipage}
\begin{minipage}{0.48\columnwidth}
c)\includegraphics[width=0.99\textwidth]{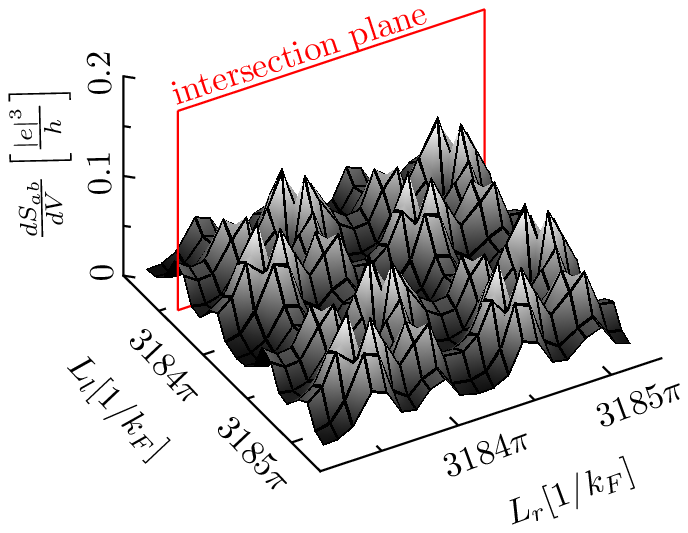}
d)\includegraphics[width=0.95\textwidth]{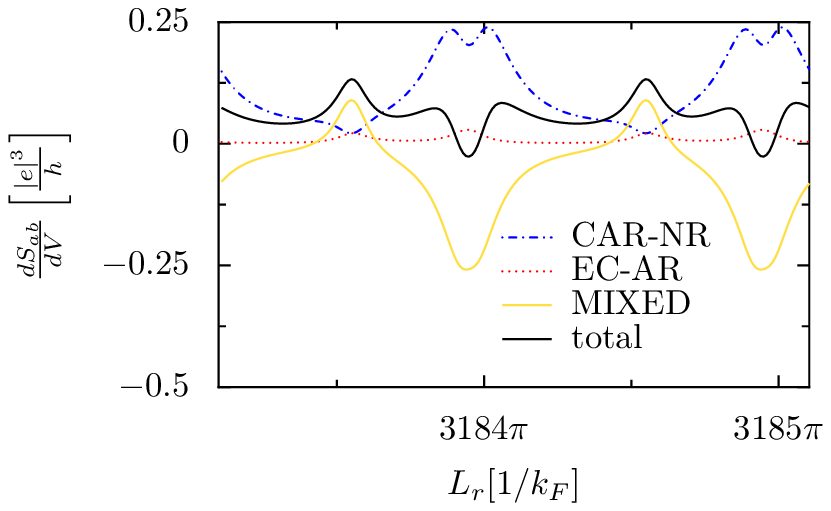}\\
\end{minipage}
\caption{\label{FP:fig:DiffSabVaVbR1T0p5}Differential current cross-correlations for symmetrically applied bias $V=V_a=V_b$ for intermediate interface transparency $T=T_{lnn}=T_{lns}=T_{rsn}=T_{rnn}=0.5$ and $R=\xi$, \newline 
a) as a function of $L_r$ and $L_l$ at low voltage $|e|V=1\cdot10^{-4}\Delta$,\newline  
b) cut through a) along one of the mountain ranges (at ${L_l=3183.7\pi/k_F}$) at higher resolution,\newline 
c) as a function of $L_r$ and $L_l$ at higher voltage $|e|V=6.25\cdot10^{-2}\Delta$,\newline
d) cut through c) at $L_l=3183.5\pi/k_F$ at higher resolution,\newline 
}
\end{figure}
\begin{figure}
\begin{minipage}{0.48\columnwidth}
a)\includegraphics[width=0.99\textwidth]{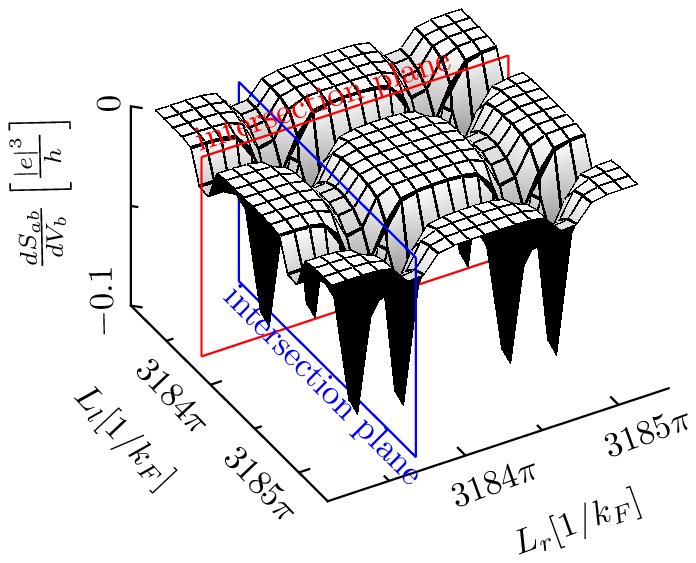}
b)\includegraphics[width=0.95\textwidth]{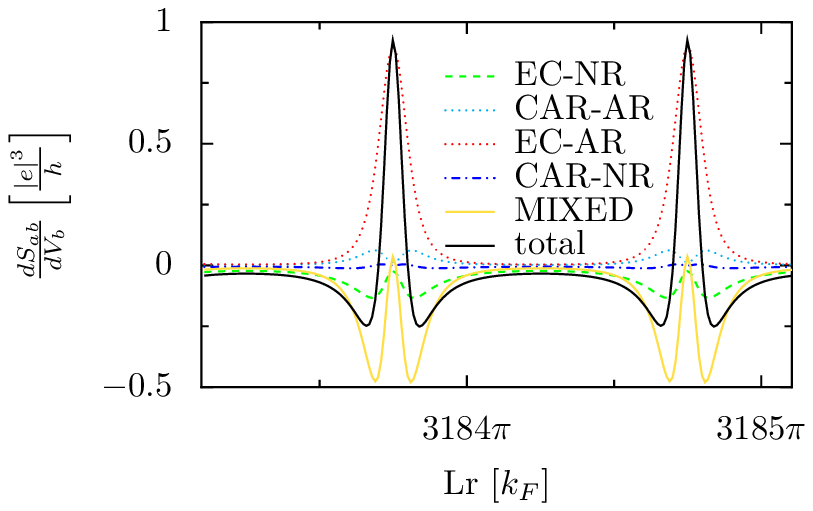}
c)\includegraphics[width=0.95\textwidth]{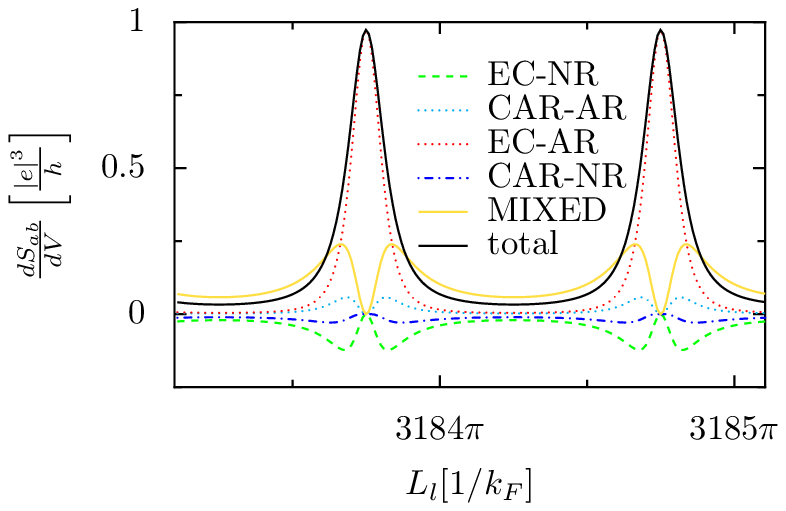}
\end{minipage}
\begin{minipage}{0.48\columnwidth}
d)\includegraphics[width=0.99\textwidth]{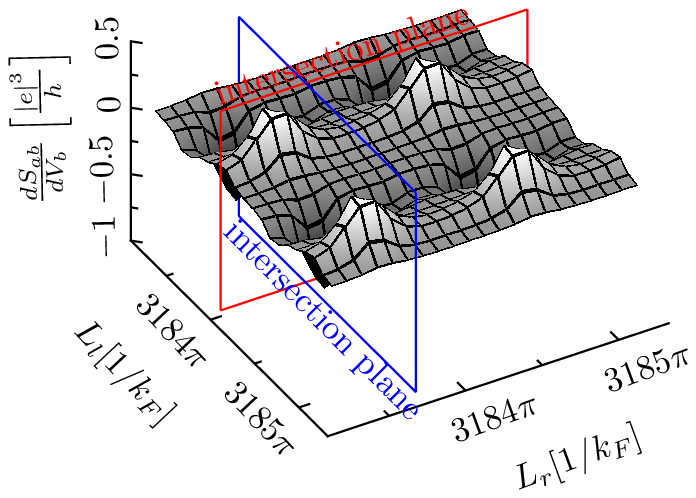}
e)\includegraphics[width=0.95\textwidth]{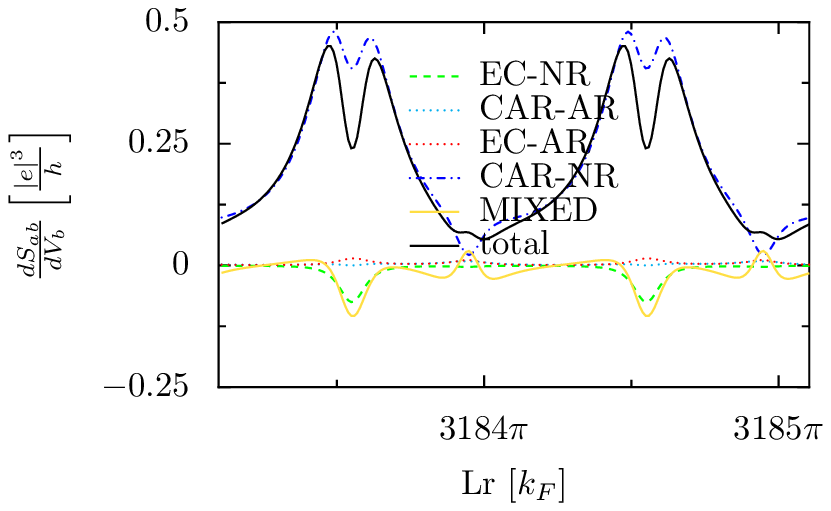}\\
f)\includegraphics[width=0.95\textwidth]{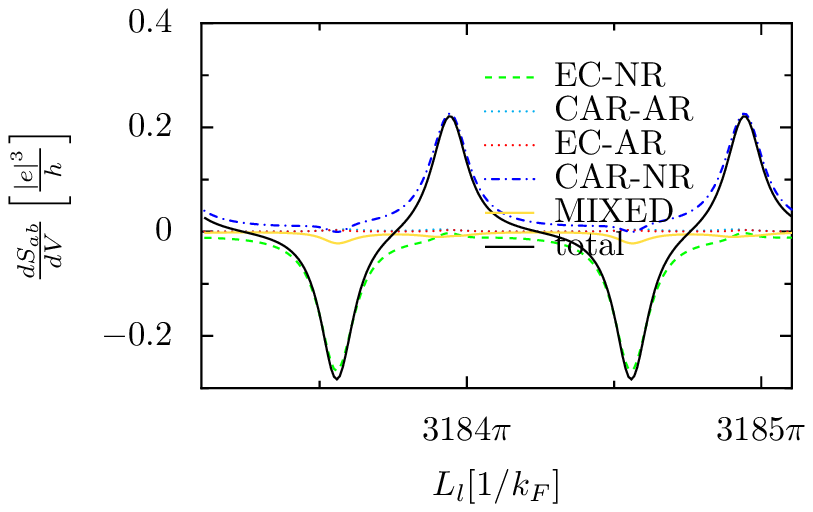}
\end{minipage}
\caption{\label{FP:fig:NoiseVbR1T0p5}Differential current cross-correlations for $V_a=0$ for intermediate interface transparency $T=T_{lnn}=T_{lns}=T_{rsn}=T_{rnn}=0.5$ and $R=\xi$, \newline 
a) as a function of $L_r$ and $L_l$ at low voltage $|e|V_b=1\cdot10^{-4}\Delta$,\newline  
b) cut through a) at ${L_l=3183.7\pi/k_F}$ at higher resolution,\newline 
c) cut through a) at ${L_r=3183.7\pi/k_F}$ at higher resolution,\newline 
c) as a function of $L_r$ and $L_l$ at higher voltage $|e|V_b=6.25\cdot10^{-2}\Delta$,\newline
d1) cut through c) at $L_l=3183.5\pi/k_F$ at higher resolution,\newline 
d2) cut through c) at $L_r=3183.7\pi/k_F$ at higher resolution.
}
\end{figure}
\subsubsection{Nonlocal conductance for $\mathbf{Va=0}$, $\mathbf{V_b>0}$ \label{paperNSN:sec:nonlocalhighT}}
Figure~\ref{FP:fig:DiffCondVbR1T0p5} shows the nonlocal conductance. The peak positions at both studied voltages have the same symmetry as  for the nonlocal conductance in the tunnel regime. But the peaks and depressions are less sharp and are higher. In the tunnel regime, the EC and the CAR peak being superimposed at low voltage were similar in magnitude. On figure~\ref{FP:fig:DiffCondVbR1T0p5}b however, the EC amplitude is much larger than the CAR amplitude. As in the symmetrically biased case, there is no CAR peak, but the forming peak features a crater. This figure illustrates that the dip in the CAR contribution occurs exactly at the peak position in the EC contribution, indicating that the competition between CAR and EC at the level of the scattering matrix is at the origin of the crater in the CAR peak.
\subsubsection{Differential current cross-correlations for a symmetrically applied bias}
Figure~\ref{FP:fig:DiffSabVaVbR1T0p5} shows the current cross-correlations for the symmetrical bias case. In the tunnel regime the current cross-correlations were dominated by CAR-NR and EC-NR as those processes depend only quadratically on the interface transparency, while MIXED has a cubic and CAR-AR, EC-AR even a bi-quadratic dependence on the interface transparency. For the current cross-correlations at intermediate interface transparency, this restriction does no longer exist. The current cross-correlations in figure~\ref{FP:fig:DiffSabVaVbR1T0p5}a feature
trenches in $L_r$ and in $L_l$ direction. Their spatial extension lets us conclude that they are connected to the AR amplitude which is independent of $L_r$ respectively $L_l$. From figure~\ref{FP:fig:DiffSabVaVbR1T0p5}b, we see that the trenches are due to MIXED. Figure~\ref{FP:fig:DiffSabVaVbR1T0p5}b shows also that the EC-AR component leads to sharp peaks at the positions where $L_r=L_l\mod\pi/k_F$. These peaks are too narrow to be resolved in the surface plot~\ref{FP:fig:DiffSabVaVbR1T0p5}a. Here we have again an example of positive current cross-correlations which are not due to CAR. The CAR-NR component which was dominant in the tunnel regime is almost zero here, which is consistent with the small amplitude of the CAR process observed in the differential conductance.

At higher energies (figures~\ref{FP:fig:DiffSabVaVbR1T0p5}d), the CAR-NR contribution gains importance as the scattering matrix elements for CAR and EC are no longer in resonance at the same position. In contrast to the tunnel regime, and like in the case of lower energies, the MIXED processes play an important role.
\subsubsection{Differential current cross-correlations for $\mathbf{Va=0}$, $\mathbf{V_b>0}$}
The current cross-correlations in the nonlocal conductance bias configuration shown in figure~\ref{FP:fig:NoiseVbR1T0p5} feature at low energy, like the ones in the symmetrically biased case, trenches where the AR parts of the scattering matrix are in resonance. The MIXED component of the current cross-correlations plays an important role and leads together with the EC-NR component to mainly negative current cross-correlations. At the points where $L_r=L_l\!\!\mod\pi/k_F$ and the EC elements of the scattering matrix are in resonance, the EC-AR component leads to positive peaks. At higher applied voltage, there is again the separation of the $L_l$ values where CAR or EC is in resonance, and MIXED becomes less important.

\subsubsection{Discussion}
The above results show that it is not necessary to be in the tunneling limit in order to separate the different components of the conductance. Against the intuitive trend, the distance in the superconductor should not be too short. Owing to multiple scattering at the different interfaces, this can be understood  as an optimum in terms of the CAR transmission, with respect to the normal barrier transmissions. In this case, the width and the height of the resonances in the conductance are wider than in the tunnel regime, which could even facilitate their observation. As far as the current cross-correlations are concerned, they are richer but also more difficult to interpret in the regime of intermediate interface transparencies because more components play a role.
%
\section{Could these Effects Be Observed Experimentally?\label{NSNpaper:sec:ExpObs}}
What are the possibilities and obstacles for an experimental observation of the phenomena discussed in this chapter?
The Fabry-Perot effect needs ballistic transport in a system with only a few (one) channels. 
The authors of the pioneering paper Ref.~\onlinecite{liang2001fabry} observed Fabry-Perot interferences in single-walled nanotubes. Instead of varying the length of the nanotube, they change the Fermi level position in the nanotube by tuning the voltage applied to a gate under the nanotube. As the resonance position depends on the product of wavenumber and length, changing the Fermi wavelength has the same effect as changing  the distance between barriers. 
This strategy is more realistic than to vary the distances $L_l$ and $L_r$. However, if the electrons in the nanotube are sensitive to the electrostatic potential of a gate, they cannot be perfectly screened. One would have to find a regime where the Coulomb interaction is small enough to be neglected in comparison to the, at least in the tunnel regime, also very small coupling energy between the nanotube and the reservoir, but high enough that the Fermi wave vector can be influenced by a gate.  

One could imagine a device consisting of two normal conducting and a superconducting electrode connected by one long carbon nanotube with a symmetry similar to devices with nanowires used in recent experiments~\cite{hofstetter2009,herrmann,hofstetter2011,das2012} where the normal and superconducting electrodes are connected over quantum dots. But while there, the system was operated in the Coulomb blockade regime, we need the other extreme of very small Coulomb energy.
On both sides of the superconductor, there has to be a gate that can be separately tuned so that
several periods of Fabry-Perot interferences can be swept in order to obtain plots similar to the one presented in the last section.

The next question concerns the interfaces. In Ref.~\onlinecite{liang2001fabry}, where Fabry-Perot interferences are observed, the nanotube is connected with near-perfect ohmic contacts to the electrodes. This implies a regime of high interface transparencies ($T\approx1$), not a tunnel regime. If the interface transparency is high, the requirement for the charging energy to be much smaller than the coupling energy is easier to meet, but in the extreme case of $T\approx1$ the contrast is too small. In the example of Ref.\onlinecite{liang2001fabry} the minimal measured conductance is $\measure{I}/\measure{V}\approx3.1e^2/h$ and the maximal measured one $\measure{I}/\measure{V}\approx3.3e^2/h$. The last section showed that it is not necessary to be in the tunnel regime to obtain a separation of the different contributions to the conductance. This can give hope that an intermediate value of interface transparency can be found. The next problem will then be to achieve this interface transparency experimentally. In addition, for sharp Fabry-Perot interferences, the length of the nanotube has to be well defined. If one uses only one long nanotube the superconducting region is created by the proximity effect in the nanotube. So one cannot really speak of a sharp interface.

In conclusion, it will be challenging to construct a device where the separation of the different components of the conductance and of the current cross-correlations by interference effects in absence of electron-electron interaction are observable. It could nevertheless be possible to reach an experimental regime where Fabry-Perot interference effects are in competition with Coulomb blockade effects. The present analysis of a simplified model describing an extreme case could help to explain such experiments in an intermediate regime and it is a good basis for a more complete theory.   
\section{Conclusion}
We have shown that positive current cross-correlations in the tunnel regime cannot be enhanced with additional barriers by a process similar to reflectionless tunneling, if an average (here, over the interbarrier lengths) has to be performed, mimicking disorder landscapes which are uncorrelated on the two sides of the set-up. In analogy, one cannot therefore expect that the use of diffusive normal metals will facilitate the experimental observation of positive current cross-correlations. This negative conclusion is important for experiments.

On the contrary, the insertion of additional tunnel barriers into ballistic NSN-devices leads to Fabry-Perot like oscillations of the differential conductance and of the current cross-correlations. The resonances for different processes are well separated at finite energy, because of the different energy dependence of electron and hole wave vectors and CAR and EC processes can be efficiently filtered for intermediate transparencies and for a central superconductor width of the order of the coherence length. 
\section*{Acknowledgements}
The authors have benefited from several fruitful discussions with B. Dou\c{c}ot. Part of this work was supported by ANR Project "Elec-EPR".
\appendix 
\section{Details on the scattering approach\label{paperNSN:app:BTK}}
The elements $s_{ij}^{\alpha\beta}$ of the scattering matrix are calculated
within the BTK approach\cite{BTK1982}: Two-component wavefunctions, where the upper component describes
electrons and the lower components holes, are matched at the interfaces and the coefficients of the resulting system of equations give the elements of the scattering matrix.\\ 
In the normal conductors the the wavefunctions are plane waves with wavevectors close to the Fermi wavevector $\hbar k_F=\sqrt{2m\mu}$. The wavevector for electrons reads $\hbar q^{+}=\sqrt{2m}\sqrt{\mu+E}$, the one for holes $\hbar q^{+}=\sqrt{2m}\sqrt{\mu-E}$. In the superconductor, the wavefunction has to obey the Bogoliubov-De Gennes equation, where the superconducting gap is supposed to be a positive constant inside the superconductor and zero outside of it. This is achieved by modifying the amplitudes of the wavefunction in the superconductor with the coherence factors $u_{E}$ and $v_{E}$ which read for energies smaller than the gap ($E<\Delta$):
\begin{align}
 u_{E}=\frac{1}{\sqrt{2}}\sqrt{1+\frac{i\sqrt{\Delta^2-E^2}}{E}},\phantom{000} 
v_{E}=\frac{1}{\sqrt{2}}\sqrt{1-\frac{i\sqrt{\Delta^2-E^2}}{E}}
\end{align}
and by using for quasi-particles proportional to 
$\bp u_E\\v_E\ep$ the wavevector $\hbar k^{+}=\sqrt{2m}\sqrt{\mu+i\sqrt{\Delta^2-E^2}}$ and for 
quasi-particles proportional to $\bp v_E\\u_E\ep$ the wavevector $\hbar k^{-}=\sqrt{2m}\sqrt{\mu-i\sqrt{\Delta^2-E^2}}$.\\
For example, the wavefunctions for an electron incoming from electrode N$_a$
take the form
\begin{widetext}
\begin{align}
\displaybreak[0]
\psi_{\mathrm{N}_a}(x)&=
\begin{pmatrix}1\\0\end{pmatrix}
\left(1\phantom{0}e^{iq^{+}x}+s^{ee}_{aa}\phantom{0} e^{-iq^{+}x}\right)
+ \begin{pmatrix}0\\1\end{pmatrix} 
\left(s^{he}_{aa}\phantom{0}e^{iq^{-}x}+0\phantom{0}e^{-iq^{-}x}\right),\\
\displaybreak[0]
\psi_{\mathrm{N}_l}(x)&=
\begin{pmatrix}1\\0\end{pmatrix}\left(c_1\phantom{0}e^{iq^{+}x}
+c_2\phantom{0}e^{-iq^{+}(x-L_l)}\right)+\begin{pmatrix}0\\1\end{pmatrix}\left(c_3\phantom{0} e^{iq^{-}x}
+c_4\phantom{0} e^{-iq^{-}(x-L_l)}\right),\\
\displaybreak[0]
\psi_{\mathrm{S}}(x)&=
\begin{pmatrix}u_E\\v_E \end{pmatrix} \left( c_5\phantom{0} e^{ik^{+}(x-L_l)}
+c_6\phantom{0} e^{-ik^{+}(x-L_l-R)}\right)\notag\\
&+\begin{pmatrix}v_E\\u_E \end{pmatrix} \left( c_7\phantom{0} e^{-ik^{-}(x-L_l}
+c_8\phantom{0} e^{ik^{-}(x-L_l-R)}\right),\\
\psi_{\mathrm{N}_r}(x)&=
\begin{pmatrix}1\\0\end{pmatrix}\left(c_9\phantom{0}e^{iq^{+}(x-L_l-R)}
+c_{10}\phantom{0}e^{-iq^{+}(x-L_l-R-L_r)}\right)\notag\\
&+\begin{pmatrix}0\\1\end{pmatrix}\left(c_{11}\phantom{0} e^{iq^{-}(x-L_l-R)}
+c_{12}\phantom{0} e^{-iq^{-}(x-L_l-R-L_r)}\right),\\
\displaybreak[0]
\psi_{\mathrm{N}_b}(x)&=
\begin{pmatrix}1\\0\end{pmatrix}\left(s^{e,e}_{b,a}\phantom{0}e^{iq^{+}(x-L_l-R-L_r)}
+0\phantom{0}e^{-iq^{+}(x-L_l-R-L_r)}\right)\notag\\
&+\begin{pmatrix}0\\1\end{pmatrix}\left(0\phantom{0} e^{iq^{-}(x-L_l-R)}
+s^{h,e}_{b,a}\phantom{0} e^{-iq^{-}(x-L_l-R-L_r)}\right)
\end{align}
\end{widetext}
in the sections N$_a$, N$_l$, S, N$_r$, and N$_b$ respectively [see
Fig.~\ref{paperNNSNN:fig:model}] and give access to the scattering matrix elements $s^{e,e}_{a,a}$, $s^{h,e}_{a,a}$, $s^{h,e}_{b,a}$ and
$s^{e,e}_{b,a}$. The remaining elements of the scattering matrix can be obtained from the other
possible scattering processes i.e. a hole incoming from electrode N$_a$, an electron/hole incoming from electrode N$_b$. 

 The interfaces are modeled by $\delta$-potentials $V(x)=Z\hbar v_F\delta(x)$, where the BTK parameter $Z$ is connected to the interface transparency $T$ by $T=(1+Z^2)^{-1}$. 
The elements of the scattering matrix can be determined and the constants $c_i$ eliminated using the continuity of the wavefunctions at
the interfaces
[$\psi_{N_a}(0)=\psi_{N_l}(0)$ etc.] and the boundary condition for
the derivatives [$\psi_{N_l}'(0)-\psi_{N_a}'(0) = Z\hbar v_F \psi_a(0)$ etc.] at every interface.\\

In simple cases, i. e. for only two or three sections and in the limit of zero energy, 
the system of equations giving the scattering matrix elements can be solved analytically (see \cite{FFM2010}), 
but in the present case of five section the expressions become so unhandy that we decided to solve the the equations numerically.
\section{Components of Current Cross Correlations\label{paperNNSNN:app:Noise}}
\begin{widetext}
Components of the current cross-correlations:
\tiny
\begin{align*}
&S_{ab}(T=0,V_a,V_b)=\frac{2e^2}{h}\int dE \phantom{0}\biggr(&&\\
&\left.
\begin{aligned}
&  2\Re\left[s_{ab}^{ee} s_{ba}^{ee} s_{aa}^{ee\dagger} s_{bb}^{ee\dagger}\right](\theta(|e|V_a-E)-2\theta(|e|V_a-E)\theta(|e|V_b-E)+\theta(|e|V_b-E))\\
+& 2\Re \left[s_{ab}^{hh} s_{ba}^{hh} s_{aa}^{hh\dagger} s_{bb}^{hh\dagger} \right](\theta(-|e|V_a-E)-2\theta(-|e|V_a-E)\theta(-|e|V_b-E)+\theta(-|e|V_b-E))
\end{aligned}\right\}{EC-NR}&&\displaybreak[0]\\ 
&\left.
\begin{aligned}
+& 2\Re \left[ s_{ba}^{eh} s_{ab}^{he} s_{aa}^{hh\dagger} s_{bb}^{ee\dagger}\right] (-\theta(-|e|V_a-E)+2\theta(-|e|V_a-E)\theta(|e|V_b-E)-\theta(|e|V_b-E))\\
+& 2\Re \left[ s_{ab}^{eh} s_{ba}^{he} s_{aa}^{ee\dagger} s_{bb}^{hh\dagger}\right] (-\theta(|e|V_a-E)+2\theta(|e|V_a-E)\theta(-|e|V_b-E)-\theta(-|e|V_b-E))\\
\end{aligned}\right\}{CAR-NR}&&\displaybreak[0]\\
&\left.
\begin{aligned}
+&  2\Re \left[s_{ab}^{hh} s_{ba}^{ee} s_{bb}^{eh\dagger} s_{aa}^{he\dagger} \right] (-\theta(|e|V_a-E)+2\theta(|e|V_a-E)\theta(-|e|V_b-E)-\theta(-|e|V_b-E))\\
+&  2\Re \left[s_{ab}^{ee} s_{ba}^{hh} s_{aa}^{eh\dagger} s_{bb}^{he\dagger} \right] (-\theta(-|e|V_a-E)+2\theta(-|e|V_a-E)\theta(|e|V_b-E)-\theta(|e|V_b-E))\\
\end{aligned}\right\}{EC-AR}&&\displaybreak[0]\\
&\left.
\begin{aligned}
+&  2\Re \left[s_{ba}^{he} s_{ab}^{he} s_{aa}^{he\dagger} s_{bb}^{he\dagger}\right] (\theta(|e|V_a-E)-2\theta(|e|V_a-E)\theta(|e|V_b-E)+\theta(|e|V_b-E))\\
+&  2\Re \left[s_{ab}^{eh} s_{ba}^{eh} s_{aa}^{eh\dagger} s_{bb}^{eh\dagger}\right] (\theta(-|e|V_a-E)-2\theta(-|e|V_a-E)\theta(-|e|V_b-E)+\theta(-|e|V_b-E))\\
\end{aligned}\right\}{CAR-AR}&&\displaybreak[0]\\ 
&\left.
\begin{aligned}
+&  2\Re \left[s_{ab}^{eh} s_{ba}^{hh} s_{bb}^{hh\dagger} s_{aa}^{eh\dagger}+ s_{ab}^{hh} s_{ba}^{eh} s_{aa}^{hh\dagger} s_{bb}^{eh\dagger}\right] (-\theta(-|e|V_a-E)+2\theta(-|e|V_a-E)\theta(-|e|V_b-E)-\theta(-|e|V_b-E))\\
+&  2\Re \left[s_{ab}^{ee} s_{ba}^{he} s_{aa}^{ee\dagger} s_{bb}^{he\dagger}+ s_{ba}^{ee} s_{ab}^{he} s_{bb}^{ee\dagger} s_{aa}^{he\dagger}\right] (-\theta(|e|V_a-E)+2\theta(|e|V_a-E)\theta(|e|V_b-E)-\theta(|e|V_b-E))\\
\end{aligned}\right\}{MIXED1} &&\displaybreak[0]\\ 
&\left.
\begin{aligned}
+&  2\Re \left[s_{ab}^{ee} s_{ba}^{eh} s_{bb}^{ee\dagger} s_{aa}^{eh\dagger}+s_{ba}^{hh} s_{ab}^{he} s_{aa}^{hh\dagger} s_{bb}^{he\dagger} \right] (\theta(-|e|V_a-E)-2\theta(-|e|V_a-E)\theta(|e|V_b-E)+\theta(|e|V_b-E))\\
+&  2\Re \left[s_{ab}^{eh} s_{ba}^{ee} s_{aa}^{ee\dagger} s_{bb}^{eh\dagger}+s_{ab}^{hh} s_{ba}^{he} s_{bb}^{hh\dagger} s_{aa}^{he\dagger} \right] (\theta(|e|V_a-E)-2\theta(|e|V_a-E)\theta(-|e|V_b-E)+\theta(-|e|V_b-E))\\
\end{aligned}\right\}{MIXED2} &&\displaybreak[0]\\ 
&\left.
\begin{aligned}
+&  2\Re \left[s_{aa}^{hh} s_{ba}^{ee} s_{ba}^{eh\dagger} s_{aa}^{he\dagger}+s_{aa}^{ee} s_{ba}^{hh} s_{ba}^{he\dagger} s_{aa}^{eh\dagger} \right] (-\theta(|e|V_a-E)+2\theta(|e|V_a-E)\theta(-|e|V_a-E)-\theta(-|e|V_a-E))\\
\end{aligned}\right\}{MIXED3a} &&\displaybreak[0]\\ 
&\left.
\begin{aligned}
+&  2\Re \left[s_{ab}^{eh} s_{bb}^{he} s_{ab}^{ee\dagger} s_{bb}^{hh\dagger}+s_{ab}^{hh} s_{bb}^{ee} s_{ab}^{he\dagger} s_{bb}^{eh\dagger}\right] (-\theta(|e|V_b-E)+2\theta(|e|V_b-E)\theta(-|e|V_b-E)-\theta(-|e|V_b-E))\\ 
\end{aligned}\right\}{MIXED3b} &&\displaybreak[0]\\ 
&\left.
\begin{aligned}
+&  2\Re \left[s_{aa}^{eh} s_{ba}^{ee} s_{aa}^{ee\dagger} s_{ba}^{eh\dagger}+s_{ba}^{hh} s_{aa}^{he} s_{aa}^{hh\dagger}s_{ba}^{he\dagger} \right] (\theta(|e|V_a-E)-2\theta(|e|V_a-E)\theta(-|e|V_a-E)+\theta(-|e|V_a-E))\\
\end{aligned}\right\}{MIXED4a} &&\displaybreak[0]\\ 
&\left.
\begin{aligned}
+&  2\Re \left[s_{ab}^{ee} s_{ab}^{eh} s_{bb}^{ee\dagger} s_{ab}^{eh\dagger})+s_{ab}^{eh} s_{bb}^{he} s_{bb}^{hh\dagger} s_{ab}^{he\dagger})\right] (\theta(|e|V_b-E)-2\theta(|e|V_b-E)\theta(-|e|V_b-E)+\theta(-|e|V_b-E))\biggl)\\
\end{aligned}\right\}{MIXED4b} &&
\end{align*}
\normalsize
Differential current cross-correlations in the nonlocal conductance setup:
\tiny
\begin{align*}
&\frac{\measure S_{ab}(T=0,V_a=0,V_b)}{\measure V_b}=\left.\right.\frac{2|e|^3}{h}\sgn(V_b)\biggl(\\
&\left.
\begin{aligned}
&\phantom{+}  2\Re\left[s_{ab}^{ee}(|e|V_b) s_{ba}^{ee}(|e|V_b) s_{aa}^{ee\dagger}(|e|V_b) s_{bb}^{ee\dagger}(|e|V_b)\right]
+ 2\Re \left[s_{ab}^{hh}(-|e|V_b) s_{ba}^{hh}(-|e|V_b) s_{aa}^{hh\dagger}(-|e|V_b) s_{bb}^{hh\dagger}(-|e|V_b)\right]\\
\end{aligned}\right\}{EC-NR}\displaybreak[0]\\ 
&\left.
\begin{aligned}
-& 2\Re \left[ s_{ba}^{eh}(|e|V_b) s_{ab}^{he}(|e|V_b) s_{aa}^{hh\dagger}(|e|V_b) s_{bb}^{ee\dagger}(|e|V_b)\right]
-  2\Re \left[ s_{ab}^{eh}(-|e|V_b) s_{ba}^{he}(-|e|V_b) s_{aa}^{ee\dagger}(-|e|V_b) s_{bb}^{hh\dagger}(-|e|V_b)\right]\\
\end{aligned}\right\}{CAR-NR}\displaybreak[0]\\
&\left.
\begin{aligned}
-&  2\Re \left[s_{ab}^{ee}(|e|V_b) s_{ba}^{hh}(|e|V_b) s_{aa}^{eh\dagger}(|e|V_b) s_{bb}^{he\dagger} (|e|V_b)\right]
-  2\Re \left[s_{ab}^{hh}(-|e|V_b) s_{ba}^{ee}(-|e|V_b) s_{bb}^{eh\dagger}(-|e|V_b) s_{aa}^{he\dagger}(-|e|V_b) \right]\\
\end{aligned}\right\}{EC-AR}\displaybreak[0]\\
&\left.
\begin{aligned}
+&  2\Re \left[s_{ba}^{he}(|e|V_b) s_{ab}^{he}(|e|V_b) s_{aa}^{he\dagger}(|e|V_b) s_{bb}^{he\dagger}(|e|V_b)\right]
+  2\Re \left[s_{ab}^{eh} (-|e|V_b)s_{ba}^{eh} (-|e|V_b)s_{aa}^{eh\dagger}(-|e|V_b) s_{bb}^{eh\dagger}(-|e|V_b)\right]\\
\end{aligned}\right\}{CAR-AR}\displaybreak[0]\\ 
&\left.
\begin{aligned}
-&  2\Re \left[s_{ab}^{ee}(|e|V_b) s_{ba}^{he}(|e|V_b) s_{aa}^{ee\dagger}(|e|V_b) s_{bb}^{he\dagger}(|e|V_b)+ s_{ba}^{ee}(|e|V_b) s_{ab}^{he}(|e|V_b) s_{bb}^{ee\dagger}(|e|V_b) s_{aa}^{he\dagger}(|e|V_b)\right]\\
-&  2\Re \left[s_{ab}^{eh}(-|e|V_b) s_{ba}^{hh}(-|e|V_b) s_{bb}^{hh\dagger}(-|e|V_b) s_{aa}^{eh\dagger}(-|e|V_b)+ s_{ab}^{hh}(-|e|V_b) s_{ba}^{eh}(-|e|V_b) s_{aa}^{hh\dagger}(-|e|V_b) s_{bb}^{eh\dagger}(-|e|V_b)\right]\\
\end{aligned}\right\}{MIXED1} &&\displaybreak[0]\\ 
&\left.
\begin{aligned}
+&  2\Re \left[s_{ab}^{eh}(|e|V_b) s_{ba}^{ee}(|e|V_b) s_{aa}^{ee\dagger}(|e|V_b) s_{bb}^{eh\dagger}(|e|V_b)+s_{ab}^{hh}(|e|V_b) s_{ba}^{he}(|e|V_b) s_{bb}^{hh\dagger}(|e|V_b) s_{aa}^{he\dagger}(|e|V_b) \right]\\
+&  2\Re \left[s_{ab}^{ee}(-|e|V_b) s_{ba}^{eh}(-|e|V_b) s_{bb}^{ee\dagger}(-|e|V_b) s_{aa}^{eh\dagger}(-|e|V_b)+s_{ba}^{hh}(-|e|V_b) s_{ab}^{he}(-|e|V_b) s_{aa}^{hh\dagger}(-|e|V_b) s_{bb}^{he\dagger}(-|e|V_b) \right]\\
\end{aligned}\right\}{MIXED2} &&\displaybreak[0]\\ 
&\left.
\begin{aligned}
-&  2\Re \left[s_{ab}^{eh}(|e|V_b) s_{bb}^{he}(|e|V_b) s_{ab}^{ee\dagger}(|e|V_b) s_{bb}^{hh\dagger}(|e|V_b)+s_{ab}^{hh}(|e|V_b) s_{bb}^{ee}(|e|V_b) s_{ab}^{he\dagger}(|e|V_b) s_{bb}^{eh\dagger}(|e|V_b)\right]\\
-&  2\Re \left[s_{ab}^{eh}(-|e|V_b) s_{bb}^{he}(-|e|V_b) s_{ab}^{ee\dagger}(-|e|V_b) s_{bb}^{hh\dagger}(-|e|V_b)+s_{ab}^{hh}(-|e|V_b) s_{bb}^{ee}(-|e|V_b) s_{ab}^{he\dagger}(-|e|V_b) s_{bb}^{eh\dagger}(-|e|V_b)\right]\\
\end{aligned}\right\}{MIXED3b} &&\displaybreak[0]\\ 
&\left.
\begin{aligned}
+&  2\Re \left[s_{ab}^{ee}(|e|V_b) s_{ab}^{eh}(|e|V_b) s_{bb}^{ee\dagger}(|e|V_b) s_{ab}^{eh\dagger}(|e|V_b)+s_{ab}^{eh}(|e|V_b) s_{bb}^{he}(|e|V_b) s_{bb}^{hh\dagger}(|e|V_b) s_{ab}^{he\dagger}(|e|V_b)\right]\\
+&  2\Re \left[s_{ab}^{ee}(-|e|V_b) s_{ab}^{eh}(-|e|V_b) s_{bb}^{ee\dagger}(-|e|V_b) s_{ab}^{eh\dagger}(-|e|V_b)+s_{ab}^{eh}(-|e|V_b) s_{bb}^{he}(-|e|V_b) s_{bb}^{hh\dagger}(-|e|V_b) s_{ab}^{he\dagger}(-|e|V_b)\right]\biggl)\\
\end{aligned}\right\}{MIXED4b} &&
\end{align*}
\normalsize
Differential current cross-correlations in the symmetrical setup:
\tiny
\begin{align*}
&\frac{dS_{ab}(T=0,V_a+V,V_b=V)}{dV}=\frac{2|e|^3}{h}\sgn(|e|V)\phantom{0}\biggr(\\
&\left.
\begin{aligned}
-& 2\Re \left[ s_{ba}^{eh}(|e|V) s_{ab}^{he}(|e|V) s_{aa}^{hh\dagger}(|e|V) s_{bb}^{ee\dagger}(|e|V)+s_{ab}^{eh}(|e|V) s_{ba}^{he}(|e|V) s_{aa}^{ee\dagger}(|e|V) s_{bb}^{hh\dagger}(|e|V)\right]\\
-& 2\Re \left[ s_{ba}^{eh}(-|e|V) s_{ab}^{he}(-|e|V) s_{aa}^{hh\dagger}(-|e|V) s_{bb}^{ee\dagger}(-|e|V)+s_{ab}^{eh}(-|e|V) s_{ba}^{he}(-|e|V) s_{aa}^{ee\dagger}(-|e|V) s_{hh}^{bb\dagger}(-|e|V)\right]\\
\end{aligned}\right\}{CAR-NR}\displaybreak[0]\\
&\left.
\begin{aligned}
-&  2\Re \left[s_{ab}^{hh}(|e|V) s_{ba}^{ee}(|e|V) s_{bb}^{eh\dagger}(|e|V) s_{aa}^{he\dagger}(|e|V)+s_{ab}^{ee}(|e|V) s_{ba}^{hh}(|e|V) s_{aa}^{eh\dagger}(|e|V) s_{bb}^{he\dagger}(|e|V) \right]\\
-&  2\Re \left[s_{ab}^{hh}(-|e|V) s_{ba}^{ee}(-|e|V) s_{bb}^{eh\dagger}(-|e|V) s_{aa}^{he\dagger}(-|e|V)+s_{ab}^{ee}(-|e|V) s_{ba}^{hh}(-|e|V) s_{aa}^{eh\dagger}(-|e|V) s_{bb}^{he\dagger}(-|e|V) \right]\\
\end{aligned}\right\}{EC-AR}\displaybreak[0]\\
&\left.
\begin{aligned}
+&  2\Re \left[s_{ab}^{ee}(|e|V) s_{ba}^{eh}(|e|V) s_{bb}^{ee\dagger}(|e|V) s_{aa}^{eh\dagger}(|e|V)+s_{ba}^{hh}(|e|V) s_{ab}^{he}(|e|V) s_{aa}^{hh\dagger}(|e|V) s_{bb}^{he\dagger}(|e|V) \right]\\
+&  2\Re \left[s_{ab}^{ee}(-|e|V) s_{ba}^{eh}(-|e|V) s_{bb}^{ee\dagger}(-|e|V) s_{aa}^{eh\dagger}(-|e|V)+s_{ba}^{hh}(-|e|V) s_{ab}^{he}(-|e|V) s_{aa}^{hh\dagger}(-|e|V) s_{bb}^{he\dagger}(-|e|V) \right]\\
+&  2\Re \left[s_{ab}^{eh}(|e|V) s_{ba}^{ee}(|e|V) s_{aa}^{ee\dagger}(|e|V) s_{bb}^{eh\dagger}(|e|V)+s_{ab}^{hh}(|e|V) s_{ba}^{he}(|e|V) s_{bb}^{hh\dagger}(|e|V) s_{aa}^{he\dagger}(|e|V) \right]\\
+&  2\Re \left[s_{ab}^{eh}(-|e|V) s_{ba}^{ee}(-|e|V) s_{aa}^{ee\dagger}(-|e|V) s_{bb}^{eh\dagger}(-|e|V)+s_{ab}^{hh}(-|e|V) s_{ba}^{he}(-|e|V) s_{bb}^{hh\dagger}(-|e|V) s_{aa}^{he\dagger}(-|e|V) \right]\\
\end{aligned}\right\}{MIXED2} &&\displaybreak[0]\\ 
&\left.
\begin{aligned}
-&  2\Re \left[s_{aa}^{hh}(|e|V) s_{ba}^{ee}(|e|V) s_{ba}^{eh\dagger}(|e|V) s_{aa}^{he\dagger}(|e|V)+s_{aa}^{ee}(|e|V) s_{ba}^{hh}(|e|V) s_{ba}^{he\dagger}(|e|V) s_{aa}^{eh\dagger}(|e|V) \right]\\
-&  2\Re \left[s_{aa}^{hh}(-|e|V) s_{ba}^{ee}(-|e|V) s_{ba}^{eh\dagger}(-|e|V) s_{aa}^{he\dagger}(-|e|V)+s_{aa}^{ee}(-|e|V) s_{ba}^{hh}(-|e|V) s_{ba}^{he\dagger}(-|e|V) s_{aa}^{eh\dagger}(-|e|V)\right]\\
\end{aligned}\right\}{MIXED3a} &&\displaybreak[0]\\ 
&\left.
\begin{aligned}
-&  2\Re \left[s_{ab}^{eh}(|e|V) s_{bb}^{he}(|e|V) s_{ab}^{ee\dagger}(|e|V) s_{bb}^{hh\dagger}(|e|V)+s_{ab}^{hh}(|e|V) s_{bb}^{ee}(|e|V) s_{ab}^{he\dagger}(|e|V) s_{bb}^{eh\dagger}(|e|V)\right]\\
-&  2\Re \left[s_{ab}^{eh}(-|e|V) s_{bb}^{he}(-|e|V) s_{ab}^{ee\dagger}(-|e|V) s_{bb}^{hh\dagger}(-|e|V)+s_{ab}^{hh}(-|e|V) s_{bb}^{ee}(-|e|V) s_{ab}^{he\dagger}(-|e|V) s_{bb}^{eh\dagger}(-|e|V)\right]\\
\end{aligned}\right\}{MIXED3b} &&\displaybreak[0]\\ 
&\left.
\begin{aligned}
+&  2\Re \left[s_{aa}^{eh}(|e|V) s_{ba}^{ee}(|e|V) s_{aa}^{ee\dagger}(|e|V) s_{ba}^{eh\dagger}(|e|V)+s_{ba}^{hh}(|e|V) s_{aa}^{he}(|e|V) s_{aa}^{hh\dagger}(|e|V)s_{ba}^{he\dagger}(|e|V) \right]\\
+&  2\Re \left[s_{aa}^{eh}(-|e|V) s_{ba}^{ee}(-|e|V) s_{aa}^{ee\dagger}(-|e|V) s_{ba}^{eh\dagger}(-|e|V)+s_{ba}^{hh}(-|e|V) s_{aa}^{he}(-|e|V) s_{aa}^{hh\dagger}(-|e|V)s_{ba}^{he\dagger}(-|e|V)\right]\\
\end{aligned}\right\}{MIXED4a} &&\displaybreak[0]\\ 
&\left.
\begin{aligned}
+&  2\Re \left[s_{ab}^{ee}(|e|V) s_{ab}^{eh}(|e|V) s_{bb}^{ee\dagger}(|e|V) s_{ab}^{eh\dagger}(|e|V)+s_{ab}^{eh}(|e|V) s_{bb}^{he}(|e|V) s_{bb}^{hh\dagger}(|e|V) s_{ab}^{he\dagger}(|e|V)\right]\\
+&  2\Re \left[s_{ab}^{ee}(-|e|V) s_{ab}^{eh}(-|e|V) s_{bb}^{ee\dagger}(-|e|V) s_{ab}^{eh\dagger}(-|e|V)+s_{ab}^{eh}(-|e|V) s_{bb}^{he}(-|e|V) s_{bb}^{hh\dagger}(-|e|V) s_{ab}^{he\dagger}(-|e|V)\right]\biggl)\\
\end{aligned}\right\}{MIXED4b}
\end{align*}
\normalsize
\end{widetext}
\section{Relations between the noise classification in the BTK and in the
Green's functions approach}
\label{PaperNSN:app:Green}
The elements of the scattering-matrix are connected to the
retarded Green's functions of the tight binding model studied in \cite{FFM2010}
via
\begin{align}
 s_{ij}^{\alpha\beta}=i\delta_{ij}+2\pi t_i t_j
\sqrt{\rho_i^{\alpha}}\sqrt{\rho_j^{\beta}}G_{ij\alpha\beta}^R
\end{align}
where $t_i$ is the transmission coefficient of the barrier $i$,
$\rho_i^{\alpha}$ the density of electron or hole states of electrode $i$ and
$G_{ij\alpha\beta}^R$ the Green's function connecting the first site in the
superconductor next to the electrode $j$ to the first site in the superconductor
next to the electrode $i$.\\
Table~\ref{PaperNSN:tab:correspondence} shows the correspondences between the
categories in the language of Green's functions and in the language of
scattering matrix elements.
\begin{table}
\begin{tabular}{l|l}
Scattering matrix& Green's function\\
classification& classification\\
\hline
\hline
CAR-NR & CAR\\
\hline
EC-AR & \ensuremath {\mathrm {AR}\text{-}\overline{\mathrm{AR}}}\\
\hline
MIXED1, MIXED2 & PRIME\\
\hline
EC-NR &EC\\
\hline
CAR-AR & AR-AR\\
\hline
MIXED3, MIXED4 & MIXED
\end{tabular}
\caption{\label{PaperNSN:tab:correspondence} Correspondences between the
categories in the language of Green's functions from~\cite{FFM2010} and in the language of
scattering matrix elements.}
\end{table}

\bibliographystyle{unsrt}
\bibliography{biblioNSN}

\begin{thebibliography}{99}
\bibitem{recher} M.~S. Choi, C. Bruder, and D. Loss, Phys. Rev. B {\bf 62}, 13569 (2000); P. Recher, E.~V. Sukhorukov, and D. Loss, Phys. Rev. B {\bf 63},  165314  (2001).
\bibitem{lesovikmartin} G.B. Lesovik, T. Martin, and G. Blatter, Eur. Phys. J. B {\bf 24},  287  (2001); N.~M. Chtchelkatchev, G. Blatter, G.~B. Lesovik, and T. Martin, Phys. Rev. B {\bf 66},
   161320  (2002).
\bibitem{allsopp} N.~K. Allsopp, V.~C. Hui, C.~J. Lambert, and S.~J. Robinson, J. Phys.: Condens. Matter {\bf 6}, 10475 (1994). 
\bibitem{byers} J.~M. Byers and M.~E. Flatt\'e, Phys. Rev. Lett. {\bf 74}, 306 (1995).
\bibitem{torres} J. Torr\`es and T. Martin, Eur. Phys. J. B {\bf 12}, 319 (1999).
\bibitem{deutscher} G. Deutscher and D. Feinberg, Appl. Phys. Lett. {\bf 76},  487 (2000).
\bibitem{falci} G. Falci, D. Feinberg, and F.~W.~J. Hekking, Europhys. Lett. {\bf 54},  255  (2001).
\bibitem{melin1} R. M\'elin and D. Feinberg, Eur. Phys. J. B {\bf 26}, 101 (2002).
\bibitem{melin2} R. M\'elin and D. Feinberg, Phys. Rev. B {\bf 70},  174509  (2004).
\bibitem{antibunching} M. B\"uttiker, Phys. Rev. B {\bf 46}, 12485 (1992).
\bibitem{bignon} G. Bignon, M. Houzet, F. Pistolesi, and F.~W.~J. Hekking, Europhys. Lett. {\bf 67},  110  (2004).
\bibitem{fazio} F.Taddei and R. Fazio, Phys. Rev. B {\bf 65}, 075317 (2002).
\bibitem{sanchez} D. S\'anchez, R. L\'opez, P. Samuelsson, and M. B\"uttiker, Phys. Rev. B {\bf 68}, 214501 (2003).
\bibitem{DelftExpt} S. Russo, M. Kroug, T.~M. Klapwijk, and A.~F. Morpurgo, Phys. Rev. Lett. {\bf 95},  027002  (2005).
\bibitem{duhotmelin} S. Duhot and R. M\'elin, Eur. Phys. J. B {\bf 53},  257
  (2006).
\bibitem{brataas} J.~P. Morten, A. Brataas, and W. Belzig, Phys. Rev. B{\bf 74}, 214510 (2006).
\bibitem{kalenkovzaikin} M.~S. Kalenkov and A.~D. Zaikin, Phys. Rev. B {\bf 75}, 172503 (2007). 
\bibitem{FFM2010} A. Freyn, M. Fl\"oser and R. M\'elin, Phys. Rev. B  {\bf 82}, 014510 (2010). 
\bibitem{exptconductance} D. Beckmann, H.~B. Weber, and H. v.~L\"ohneysen, Phys. Rev. Lett. {\bf 93},  197003  (2004); 
P. Cadden-Zimansky and V. Chandrasekhar, Phys. Rev. Lett. {\bf 97},  237003
(2006); 
P. Cadden-Zimansky, Z. Jiang, and V. Chandrasekhar, New Journal of Physics {\bf 9},  116  (2007); 
P. Cadden-Zimansky, J. Wei and V. Chandrasekhar, Nature Physics {\bf 5}, 393 (2009); T. Noh, S. Davis, and V. Chandrasekhar, Arxiv preprint arXiv:1210.8426.
\bibitem{exptcrossnoise} J. Wei and V. Chandrasekhar, Nature Physics {\bf 6}, 494 (2010); B. Kaviraj, O. Coupiac, H. Courtois, and F. Lefloch, Phys. Rev. Lett. {\bf 107}, 077005 (2011).
\bibitem{das2012}
A. Das, Y. Ronen, M. Heiblum, D. Mahalu, A.~V. Kretinin, and H. Shtrikman, Arxiv
  preprint arXiv:1205.2455  (2012).
\bibitem{melinbenjamin} R. M\' elin, C. Benjamin, and T. Martin, Phys. Rev. B {\bf 77}, 094512 (2008). 
\bibitem{reflectionlesstunnexpt} A. Kastalsky, A.~W. Kleinsasser, L.~H. Greene, R. Bhat, F.~P. Milliken and J.~P. Harbison, Phys. Rev. Lett. {\bf 67}, 3026 (1991). 
\bibitem{reflectionlesstunntheo} F.~W.~J. Hekking and Yu.~V. Nazarov, Phys. Rev. Lett. {\bf 71}, 1625 (1991); ibid., Phys. Rev. B {\bf 49}, 6847 (1994).
\bibitem{MB1994} J.~A. Melsen and Beenakker C.~W. J., Physica B {\bf 203},  219  (1994).
\bibitem{BTK1982}
G.~E. Blonder, M. Tinkham, and T.~M. Klapwijk, Phys. Rev. B {\bf 25},  4515 (1982).
\bibitem{AD1996}
M.~P. Anantram and S. Datta, Phys. Rev. B {\bf 53},  16390  (1996).
\bibitem{ML1992} T. Martin and R. Landauer, Phys. Rev. B {\bf 45},  1742  (1992).
\bibitem{Buettiker1990} M. B\"uttiker, Phys. Rev. Lett. {\bf 65},  2901  (1990). 
\bibitem{liang2001fabry}
W. Liang, M. Bockrath, D. Bozovic,  J.~H. Hafner, M. Tinkham, H. Park, Nature {\bf 411},  665  (2001).
\bibitem{PhysRevLett.96.207003}
H.~I. J\o{}rgensen, K. Grove-Rasmussen, T. Novotn\'y, K. Flensberg, and P.~E. Lindelof, Phys. Rev. Lett. {\bf 96}, 207003 (2006).
\bibitem{saint-james} P. G. de Gennes and D. Saint-James, Phys. Lett. {\bf 4}, 151 (1963); D. Saint-James, J. Phys. (Paris) {\bf 25}, 899 (1964).
\bibitem{tomasch} W. J. Tomasch, Phys. Rev. Lett. {\bf 15}, 672 (1965); ibid. , Phys. Rev. Lett. {\bf 16}, 16 (1966); W. L. McMillan and P. W. Anderson, Phys. Rev. Lett. {\bf 16}, 85 (1966); J. M. Rowell and W. L. McMillan, Phys. Rev. Lett. {\bf 16}, 453 (1966).
\bibitem{entin-wohlmann} O. Entin-Wohlmann and J. Bar-Sagi, Phys. Rev. B {\bf 18}, 3174 (1978).
\bibitem{hofstetter2009}
L. Hofstetter, S. Csonka, J. Nyg{\aa}rd, and C. Sch{\"o}nenberger, {\em Cooper
  pair splitter realized in a two-quantum-dot Y-junction}, Nature {\bf 461},
  960  (2009).
\bibitem{herrmann}
L.~G. Herrmann, F. Portier, P. Roche, A. Levy Yeyati,  T. Kontos, and C. Strunk, Phys. Rev. Lett. {\bf 104},  026801  (2010).
\bibitem{hofstetter2011}
L. Hofstetter, S. Csonka, A. Baumgartner, G.  F\"ul\"op, S. d'Hollosy, J. Nyg\aa{}rd, and C. Sch\"onenberger, Phys. Rev.
  Lett. {\bf 107},  136801  (2011).


\end{thebibliography}

\end{document}